\documentclass[a4paper,11pt]{article}
\usepackage{jcappub} % for details on the use of the package, please see the JINST-author-manual
\usepackage{xspace}
\usepackage[modulo]{lineno}
\usepackage{graphicx}
\usepackage{txfonts}
\usepackage{fontawesome}
\usepackage{xcolor}
\usepackage{hyperref} 
\usepackage{orcidlink}
\usepackage{booktabs} 

\newcommand{\hpmpc}{$h$$\cdot$Mpc$^{-1}$}
\newcommand{\mpc}{h^{-1}\text{$\cdot$Mpc}}

\newcommand{\sne}{SNe Ia\xspace}
\newcommand{\sn}{SN Ia\xspace}
\newcommand{\fsi}{f\sigma_8}
\newcommand{\kmax}{k_{\rm max}}
\newcommand{\sigu}{\sigma_{\mathrm{u}}}
\newcommand{\sigg}{\sigma_{\mathrm{g}}}
\newcommand{\fsifid}{(f\sigma_8)_{\mathrm{fid}}}
\newcommand{\bsi}{b\sigma_8}
\newcommand{\si}{\sigma_8}
\newcommand{\lcdm}{$\Lambda\mathrm{CDM}$}

\newcommand{\git}[2]{\href{https://github.com/#1}{\faGithub}\footnote{\label{#1}#2\url{https://github.com/#1}}}
\newcommand{\irow}[1]{
  \begin{matrix}(\,#1\,)\end{matrix}
}

\arxivnumber{1234.56789}

\title{\boldmath Field-level inference of growth rate from DESI BGS galaxy and ZTF supernova simulations}

\author[1]{Corentin Ravoux\orcidlink{0000-0002-3500-6635}}
\author[2]{Bastien Carreres\orcidlink{0000-0002-7234-844X}}
\author[3]{Julian Bautista\orcidlink{0000-0002-9885-3989}}
\author[3]{Damiano Rosselli\orcidlink{0000-0001-6839-1421}}
\author[4]{Jessica Aguilar}
\author[5]{Steven Ahlen\orcidlink{0000-0001-6098-7247}}
\author[6]{Melissa Amenouche\orcidlink{0009-0006-7454-3579}}
\author[7,8]{Uendert Andrade\orcidlink{0000-0002-4118-8236}}
\author[9]{Eric Armengaud\orcidlink{0000-0001-7600-5148}}
\author[10,11]{Alejandro Aviles\orcidlink{0000-0001-5998-3986}}
\author[12]{Florian Beutler\orcidlink{0000-0003-0467-5438}}
\author[13,14]{Davide Bianchi\orcidlink{0000-0001-9712-0006}}
\author[15]{David Brooks}
\author[16]{Umut Burgaz\orcidlink{0000-0003-0126-3999}}
\author[17,18]{Aurelio Carnero Rosell\orcidlink{0000-0003-3044-5150}}
\author[4]{Edmond Chaussidon\orcidlink{0000-0001-8996-4874}}
\author[4]{Todd Claybaugh}
\author[4]{Andrei Cuceu\orcidlink{0000-0002-2169-0595}}
\author[19]{Axel de la Macorra\orcidlink{0000-0002-1769-1640}}
\author[20]{Georgios Dimitriadis\orcidlink{0000-0001-9494-179X}}
\author[21,22]{Andreu Font-Ribera\orcidlink{0000-0002-3033-7312}}
\author[23]{Daniel Felipe Forero Sánchez\orcidlink{0000-0001-5957-332X}}
\author[24,25]{Jaime E. Forero-Romero\orcidlink{0000-0002-2890-3725}}
\author[26,27,28]{Enrique Gaztañaga\orcidlink{0000-0001-9632-0815}}
\author[29]{Satya Gontcho A Gontcho\orcidlink{0000-0003-3142-233X}}
\author[30]{Ariel Goobar\orcidlink{0000-0002-4163-4996}}
\author[31]{Gaston Gutierrez}
\author[9,32]{Hiram K. Herrera-Alcantar\orcidlink{0000-0002-9136-9609}}
\author[33]{Cullan Howlett\orcidlink{0000-0002-1081-9410}}
\author[34]{Mustapha Ishak\orcidlink{0000-0002-6024-466X}}
\author[30]{Joel Johansson\orcidlink{0000-0001-5975-290X}}
\author[35]{Stephanie Juneau\orcidlink{0000-0002-0000-2394}}
\author[15]{Ofer Lahav\orcidlink{0000-0002-1134-9035}}
\author[4]{Martin Landriau\orcidlink{0000-0003-1838-8528}}
\author[36]{Laurent Le Guillou\orcidlink{0000-0001-7178-8868}}
\author[37,38]{Alexie Leauthaud\orcidlink{0000-0002-3677-3617}}
\author[22,39]{Marc Manera\orcidlink{0000-0003-4962-8934}}
\author[21,22]{Ramon Miquel}
\author[27]{Seshadri Nadathur\orcidlink{0000-0001-9070-3102}}
\author[40]{Jakob Nordin\orcidlink{0000-0001-8342-6274}}
\author[11,19]{Hernan Enrique Noriega\orcidlink{0000-0002-3397-3998}}
\author[41,42]{Enrique Paillas\orcidlink{0000-0002-4637-2868}}
\author[4,9]{Nathalie Palanque-Delabrouille\orcidlink{0000-0003-3188-784X}}
\author[43,44,45]{Will Percival\orcidlink{0000-0002-0644-5727}}
\author[46]{Francisco Prada\orcidlink{0000-0001-7145-8674}}
\author[47]{Ignasi P\'erez-R\`afols\orcidlink{0000-0001-6979-0125}}
\author[48]{Mickael Rigault\orcidlink{0000-0002-8121-2560}}
\author[49]{Antoine Rocher\orcidlink{0000-0003-4349-6424}}
\author[1]{Philippe Rosnet\orcidlink{0000-0002-6099-7565}}
\author[50]{Graziano Rossi}
\author[51]{Rossana Ruggeri\orcidlink{0000-0002-0394-0896}}
\author[52,53]{Lado Samushia\orcidlink{0000-0002-1609-5687}}
\author[54]{Eusebio Sanchez\orcidlink{0000-0002-9646-8198}}
\author[55]{Christoph Saulder\orcidlink{0000-0002-0408-5633}}
\author[4]{David Schlegel}
\author[8,56]{Michael Schubnell}
\author[57]{Hee-Jong Seo\orcidlink{0000-0002-6588-3508}}
\author[4]{Joseph Silber\orcidlink{0000-0002-3461-0320}}
\author[18,28]{Małgorzata Siudek\orcidlink{0000-0002-2949-2155}}
\author[20]{Mathew Smith\orcidlink{0000-0002-3321-1432}}
\author[8]{Gregory Tarl\'{e}\orcidlink{0000-0003-1704-0781}}
\author[35]{Benjamin A. Weaver}
\author[36]{Pauline Zarrouk\orcidlink{0000-0002-7305-9578}}

\affiliation[1]{Université Clermont-Auvergne, CNRS, LPCA, 63000 Clermont-Ferrand, France}
\affiliation[2]{Department of Physics, Duke University Durham, NC 27708, USA}
\affiliation[3]{Aix Marseille Université, CNRS/IN2P3, CPPM, Marseille, France}
\affiliation[4]{Lawrence Berkeley National Laboratory, 1 Cyclotron Road, Berkeley, CA 94720, USA}
\affiliation[5]{Department of Physics, Boston University, 590 Commonwealth Avenue, Boston, MA 02215 USA}
\affiliation[6]{Centre Spatial de Liège, Université de Liège, Avenue du Pré-Aily, 4031 Angleur, Belgium}
\affiliation[7]{Leinweber Center for Theoretical Physics, University of Michigan, 450 Church Street, Ann Arbor, Michigan 48109-1040, USA}
\affiliation[8]{University of Michigan, 500 S. State Street, Ann Arbor, MI 48109, USA}
\affiliation[9]{IRFU, CEA, Universit\'{e} Paris-Saclay, F-91191 Gif-sur-Yvette, France}
\affiliation[10]{Instituto Avanzado de Cosmolog\'{\i}a A.~C., San Marcos 11 - Atenas 202. Magdalena Contreras. Ciudad de M\'{e}xico C.~P.~10720, M\'{e}xico}
\affiliation[11]{Instituto de Ciencias F\'{\i}sicas, Universidad Nacional Aut\'onoma de M\'exico, Av. Universidad s/n, Cuernavaca, Morelos, C.~P.~62210, M\'exico}
\affiliation[12]{Institute for Astronomy, University of Edinburgh, Royal Observatory, Blackford Hill, Edinburgh EH9 3HJ, UK}
\affiliation[13]{Dipartimento di Fisica ``Aldo Pontremoli'', Universit\`a degli Studi di Milano, Via Celoria 16, I-20133 Milano, Italy}
\affiliation[14]{INAF-Osservatorio Astronomico di Brera, Via Brera 28, 20122 Milano, Italy}
\affiliation[15]{Department of Physics \& Astronomy, University College London, Gower Street, London, WC1E 6BT, UK}
\affiliation[16]{School of Physics, Trinity College Dublin, College Green, Dublin 2, Ireland}
\affiliation[17]{Departamento de Astrof\'{\i}sica, Universidad de La Laguna (ULL), E-38206, La Laguna, Tenerife, Spain}
\affiliation[18]{Instituto de Astrof\'{\i}sica de Canarias, C/ V\'{\i}a L\'{a}ctea, s/n, E-38205 La Laguna, Tenerife, Spain}
\affiliation[19]{Instituto de F\'{\i}sica, Universidad Nacional Aut\'{o}noma de M\'{e}xico,  Circuito de la Investigaci\'{o}n Cient\'{\i}fica, Ciudad Universitaria, Cd. de M\'{e}xico  C.~P.~04510,  M\'{e}xico}
\affiliation[20]{Department of Physics, Lancaster University, Lancaster, LA1 4YB, UK}
\affiliation[21]{Instituci\'{o} Catalana de Recerca i Estudis Avan\c{c}ats, Passeig de Llu\'{\i}s Companys, 23, 08010 Barcelona, Spain}
\affiliation[22]{Institut de F\'{i}sica d’Altes Energies (IFAE), The Barcelona Institute of Science and Technology, Edifici Cn, Campus UAB, 08193, Bellaterra (Barcelona), Spain}
\affiliation[23]{Institut de Ci\`encies del Cosmos (ICCUB), Universitat de Barcelona (UB), c. Mart\'i i Franqu\`es, 1, 08028 Barcelona, Spain}
\affiliation[24]{Departamento de F\'isica, Universidad de los Andes, Cra. 1 No. 18A-10, Edificio Ip, CP 111711, Bogot\'a, Colombia}
\affiliation[25]{Observatorio Astron\'omico, Universidad de los Andes, Cra. 1 No. 18A-10, Edificio H, CP 111711 Bogot\'a, Colombia}
\affiliation[26]{Institut d'Estudis Espacials de Catalunya (IEEC), c/ Esteve Terradas 1, Edifici RDIT, Campus PMT-UPC, 08860 Castelldefels, Spain}
\affiliation[27]{Institute of Cosmology and Gravitation, University of Portsmouth, Dennis Sciama Building, Portsmouth, PO1 3FX, UK}
\affiliation[28]{Institute of Space Sciences, ICE-CSIC, Campus UAB, Carrer de Can Magrans s/n, 08913 Bellaterra, Barcelona, Spain}
\affiliation[29]{University of Virginia, Department of Astronomy, Charlottesville, VA 22904, USA}
\affiliation[30]{The Oskar Klein Centre, Department of Physics, AlbaNova, SE-10691 Stockholm, Sweden}
\affiliation[31]{Fermi National Accelerator Laboratory, PO Box 500, Batavia, IL 60510, USA}
\affiliation[32]{Institut d'Astrophysique de Paris. 98 bis boulevard Arago. 75014 Paris, France}
\affiliation[33]{School of Mathematics and Physics, University of Queensland, Brisbane, QLD 4072, Australia}
\affiliation[34]{Department of Physics, The University of Texas at Dallas, 800 W. Campbell Rd., Richardson, TX 75080, USA}
\affiliation[35]{NSF NOIRLab, 950 N. Cherry Ave., Tucson, AZ 85719, USA}
\affiliation[36]{Sorbonne Universit\'{e}, CNRS/IN2P3, Laboratoire de Physique Nucl\'{e}aire et de Hautes Energies (LPNHE), FR-75005 Paris, France}
\affiliation[37]{Department of Astronomy and Astrophysics, UCO/Lick Observatory, University of California, 1156 High Street, Santa Cruz, CA 95064, USA}
\affiliation[38]{Department of Astronomy and Astrophysics, University of California, Santa Cruz, 1156 High Street, Santa Cruz, CA 95065, USA}
\affiliation[39]{Departament de F\'{i}sica, Serra H\'{u}nter, Universitat Aut\`{o}noma de Barcelona, 08193 Bellaterra (Barcelona), Spain}
\affiliation[40]{Institut für Physik, Humboldt-Universität zu Berlin, Newtonstr. 15, 12489 Berlin, Germany}
\affiliation[41]{Instituto de Estudios Astrof\'isicos, Facultad de Ingenier\'ia y Ciencias, Universidad Diego Portales, Av. Ej\'ercito Libertador 441, Santiago, Chile}
\affiliation[42]{Steward Observatory, University of Arizona, 933 N. Cherry Avenue, Tucson, AZ 85721, USA}
\affiliation[43]{Department of Physics and Astronomy, University of Waterloo, 200 University Ave W, Waterloo, ON N2L 3G1, Canada}
\affiliation[44]{Perimeter Institute for Theoretical Physics, 31 Caroline St. North, Waterloo, ON N2L 2Y5, Canada}
\affiliation[45]{Waterloo Centre for Astrophysics, University of Waterloo, 200 University Ave W, Waterloo, ON N2L 3G1, Canada}
\affiliation[46]{Instituto de Astrof\'{i}sica de Andaluc\'{i}a (CSIC), Glorieta de la Astronom\'{i}a, s/n, E-18008 Granada, Spain}
\affiliation[47]{Departament de F\'isica, EEBE, Universitat Polit\`ecnica de Catalunya, c/Eduard Maristany 10, 08930 Barcelona, Spain}
\affiliation[48]{IP2I Lyon/IN2P3, CNRS, Universite Claude Bernard Lyon 1, UMR5822, F-69622, Villeurbanne, France}
\affiliation[49]{Institute of Physics, Laboratory of Astrophysics, \'{E}cole Polytechnique F\'{e}d\'{e}rale de Lausanne (EPFL), Observatoire de Sauverny, Chemin Pegasi 51, CH-1290 Versoix, Switzerland}
\affiliation[50]{Department of Physics and Astronomy, Sejong University, 209 Neungdong-ro, Gwangjin-gu, Seoul 05006, Republic of Korea}
\affiliation[51]{Queensland University of Technology,  School of Chemistry \& Physics, George St, Brisbane 4001, Australia}
\affiliation[52]{Abastumani Astrophysical Observatory, Tbilisi, GE-0179, Georgia}
\affiliation[53]{Department of Physics, Kansas State University, 116 Cardwell Hall, Manhattan, KS 66506, USA}
\affiliation[54]{CIEMAT, Avenida Complutense 40, E-28040 Madrid, Spain}
\affiliation[55]{Max Planck Institute for Extraterrestrial Physics, Gie\ss enbachstra\ss e 1, 85748 Garching, Germany}
\affiliation[56]{Department of Physics, University of Michigan, 450 Church Street, Ann Arbor, MI 48109, USA}
\affiliation[57]{Department of Physics \& Astronomy, Ohio University, 139 University Terrace, Athens, OH 45701, USA}

\emailAdd{corentin.ravoux@clermont.in2p3.fr}

\abstract{Galaxy peculiar velocities are a powerful probe of the growth rate of structure $f$ at low redshifts ($z < 0.2$), measured through the combination $\fsi$. These constraints are improved by combining peculiar velocities with redshift space distortion measurements from the galaxy density field. We create a dedicated simulation to test growth rate measurement from the combination of Type Ia supernova peculiar velocities from the Zwicky Transient Facility (ZTF) with the Bright Galaxy Survey (BGS) of the Dark Energy Spectroscopic Instrument (DESI), using projections for 6 years of ZTF data and the Data Release 3 (DR3) of DESI. We apply a wide-angle likelihood-based field-level inference to combine ZTF velocity and DESI BGS density fields simultaneously. This procedure is performed while varying the different density and velocity field configurations. Adding DESI galaxy density information to the ZTF supernovae improves the constraint on nuisance parameters and yields up to a 50 \% reduction in the growth-rate uncertainty without introducing bias. If we consider the conservatively best DESI density configuration, the mean error bar achievable by a DESI ZTF measurement is $\sigma_{\fsi} = {}^{+0.047}_{-0.048}$ corresponding to a 10 \% error. These measurements will play a critical role for testing modified gravity theories.}

\begin{document}
\maketitle
\flushbottom

\section{Introduction}

The current \lcdm~standard model of cosmology relies on the theory of General Relativity (GR) and assumes that the Universe is composed of a cosmological constant ($\Lambda$), cold dark matter (CDM), baryonic matter, neutrinos, and photons. Although the \lcdm~model has produced many results in agreement with observations, the unknown nature of dark matter and dark energy leaves it incomplete (see, e.g., \cite{Frieman2008,Turner2022} for \lcdm~reviews). Furthermore, the increasing precision of cosmological measurements starts to challenge the \lcdm~model with the emergence of strong tensions~\cite{CosmoVerse2025}. In particular, recent measurements from the Dark Energy Spectroscopic Instrument (DESI)~\cite{DESI2024a,DESI2025,DESI2025b} showcase a strong hint for evolving dark energy, weakening the cosmological constant hypothesis.

Several theories can address the challenges of explaining the accelerated expansion. The first class of theories considers this unknown as a fluid, a scalar field, or a fifth force. The quintessence~\cite{Tsujikawa2013} or k-essence~\cite{ArmendarizPicon2000} models are among the most widely used models. The second class of theories, called modified gravity, avoids the introduction of a dark energy component by modifying GR at cosmological scales (see, e.g.~\cite{Shankaranarayanan2022} for a complete review). Widely used examples are generalized Brans–Dicke~\cite{DeFelice2010}, $f(R)$~\cite{Sotiriou2008}, DGP~\cite{Dvali2000} with its non-self-accelerating nDGP extension, or the more general form of Horndeski~\cite{Kobayashi2019} theories. Solar system tests of modified gravity at small scales~\cite{Bertotti2003,Merkowitz2010}, as well as the measurement of a gravitational wave with an electromagnetic counterpart~\cite{LIGOScientific2017}, have strongly constrained most modified theories. In particular, the remaining survivors, such as $f(R)$ or nDGP, are not sufficient to fully explain the accelerating expansion of the Universe. However, they remain interesting as benchmark models to illustrate the impact of modified gravity theories on observables, such as the ones used in this work.

An efficient way to test GR at cosmological scales is to probe the cosmic web dynamics by constraining the growth rate of large-scale structures ($f$). This term is defined in standard perturbation theory, on scales larger than the Jeans length, as the logarithmic derivative with respect to the scale factor of the growing mode of structure formation. Although strictly defined in the linear regime, $f$ is commonly used as a summary statistic for cosmic-web dynamics. The growth rate directly links large-scale densities to velocities in the cosmic web. A good approximation in GR is given by $f(z) = \Omega_m(z)^{0.55}$~\cite{linder_cosmic_2005}. The modified gravity theories cited above can give similar results regarding background cosmology while predicting very different $f(z)$ evolutions. Thus, constraints on this parameter allow us to discriminate between the modified gravity theories. Combined with observables that probe background cosmology (BAO, CMB,...), or directly by assuming GR, these constraints can be directly translated to dark energy equation of state (see, e.g.~\cite{Ishak2018,Hou2023} for a recent review). In practice, the growth rate is entirely degenerate with the amplitude of the linear power spectrum, often parameterized by $\si$, such that the combination $\fsi$~is used.

Measuring peculiar velocities (PVs) of extra-galactic objects gives constraints on $\fsi$. PVs are derived from a distance estimate based on standardizable candles such as Type Ia supernovae (\sne)~\cite{Tripp1998} or empirical relations such as Tully-Fisher (TF) \citep{tully_1977} and Fundamental plane (FP)~\citep{Djorgovski_1987}. By simultaneously measuring the observed redshift of an object, those distance estimators can be used to break the degeneracy between the Hubble flow and the object's motion. This process yields a direct measurement of the peculiar velocities of that object. Growth rate constraints are then obtained with various methods: using compressed two-point statistics such as the density and momentum correlation function, power spectra, or the average pair-wise velocities \citep{Ferreira1998,Dupuy2019,Howlett2019,Turner2022,Qin2025}, directly comparing PV to another estimation from reconstruction techniques~\citep{Springob2014,Carrick2015,Boruah2020,Said2020,Qin2023,Boubel2023}, using a likelihood-based field-level estimator~\citep{Johnson2014,howlett_measuring_2017,adams_improving_2017,huterer_testing_2017,adams_joint_2020,lai_using_2022,carreres_growth-rate_2023,Ravoux2025,laiDESIDR1Peculiar2025} or with forward modeling, or simulation-based inference~\citep{boruah_reconstructing_2021,Valade2022a,Valade2022b,Pfeifer2023}.

The $\fsi$~parameter can also be constrained by measuring the impact of redshift space distortion (RSD) on galaxy clustering statistics. With this method, the impact of PVs is measured indirectly by the deformation created in statistical measurements. Most constraints come from the two-point correlation function of galaxies or quasars, with e.g., the extended Baryon Oscillation Spectroscopic Survey (eBOSS)~\cite{Bautista2021,GilMarin2020,Tamone2020,deMattia2021,Hou2021,Neveux2020} or, more recently, with the Dark Energy Spectroscopic Instrument (DESI)~\cite{desi_collaboration_desi_2016,desi_collaboration_desi_2016-1,DESI2022,DESI2023,DESI2023b,DESI2023c,DESI2024b,DESI2024c}.

Growth rate constraints from RSD give lower error bars for redshift $z \gtrsim 0.3$ where probes are less limited by volume, whereas PV measurements, less sensitive to cosmic variance, provide better constraints at low redshift, $z \lesssim 0.2$, despite the small sample sizes. Combining these two methods over common redshift ranges gives optimal $\fsi$~constraints. In this paper, we use the likelihood-based field-level inference method developed in~\cite{Ravoux2025} with the \texttt{flip}~\git{corentinravoux/flip}{Field Level Inference Package } software to combine RSD and PV measurements on simulations. We simulate peculiar velocities of \sne~obtained from the Zwicky Transient Facility (ZTF)~\cite{graham_2019,Bellm2019,Rigault2024} to reproduce the study performed in~\cite{carreres_growth-rate_2023} (noted \textbf{C23} hereafter). We couple those PV measurements with simulations of the low-redshift bright galaxy survey (BGS) sample mimicking DESI Data Release 3 (noted DESI DR3 hereafter). We created dedicated \sn~simulations mimicking the complete ZTF survey with the \texttt{snsim}~\git{bastiencarreres/snsim}{} software on the \texttt{AbacusSummit}~\citep{Maksimova2021,Garrison2021} halo catalogs, which we tuned to reproduce the DESI BGS density field. Performing this measurement on simulations allows us to test our method on realistic datasets while controlling potentially sources of systematics and forecasting precisely the constraints achievable by combining DESI with ZTF. Note that in this work, we did not attempt to use mock versions of the DESI Peculiar Velocity survey~\cite{Bautista2025}, containing PVs from Tully-Fisher and Fundamental Plane methods, as it goes beyond the scope of this work.

This article is organized as follows. In section \ref{sec:simulations}, we detail the generation of the ZTF \sne~and DESI BGS simulations used in this study. Section~\ref{subsec:method} gives an overview of the likelihood-based field-level inference method. Section~\ref{sec:results} shows the fit results of the DESI BGS density field, the ZTF type Ia supernova (\sn)~velocity field, and the combination of both fields. Finally, section~\ref{sec:conclusion} discusses possible improvements to this study.

\section{Simulating DESI BGS galaxies and ZTF supernovae}
\label{sec:simulations}

\subsection{Abacus DESI BGS simulations}
\label{subsec:abacus}

\begin{figure*}
    \centering
	\includegraphics[trim=2.2cm 1.1cm 1cm 1.2cm, clip=true, width=0.85\textwidth]{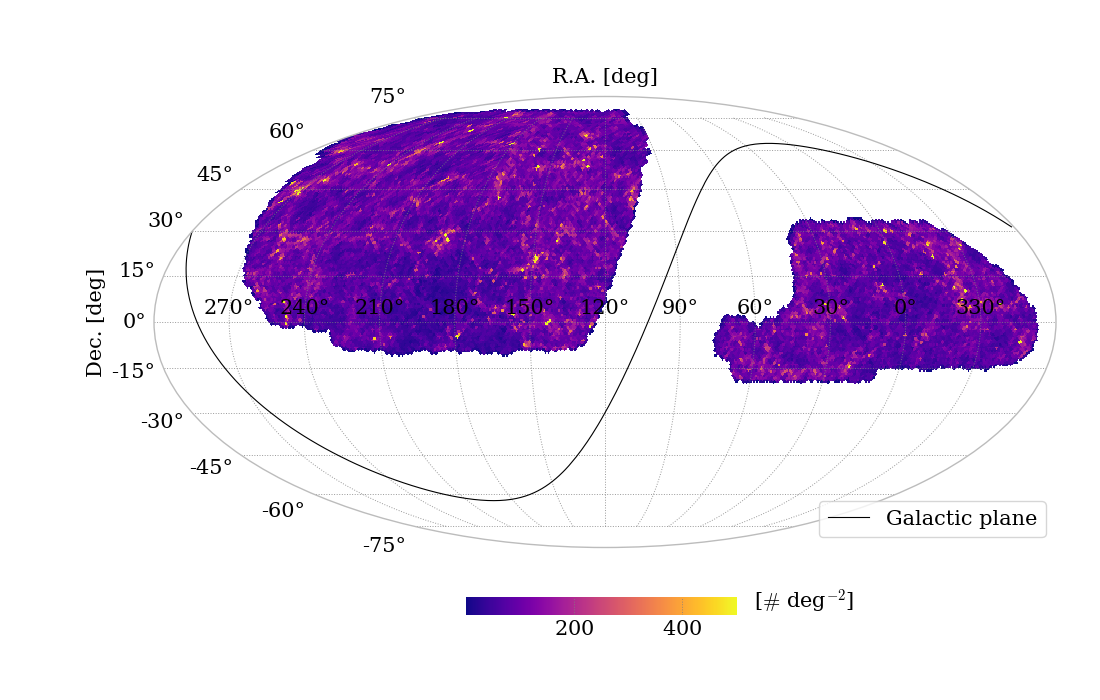}
    \caption{Mollweide representation (with resolution parameter $n_{\mathrm{side}}= 64$) of the planned DESI DR3 BGS galaxy survey for one \texttt{AbacusSummit} sub-box. The number density is normalized by the constant size of each healpix pixel. This galaxy field is expressed at the redshift of the simulation $z=0.2$ and the volume probed corresponds to a sphere around the observer with a maximal redshift $z_{\mathrm{max}} = 0.11$.}
    \label{fig:density_field}
\end{figure*}

We use the publicly available \texttt{AbacusSummit} N-body simulation \citep{Maksimova2021,Garrison2021}, and in particular, the base simulation with phase $000$ containing $6912^3$ particles in a 2 $\mathrm{Gpc}\cdot h^{-1}$ box. We only use one simulation, as the aim of this paper is to test all the fitting configurations in an extensive way. We note that this number of simulations might not be sufficient for the complete characterization of the error bar and measurement covariance; we defer this test for future studies. This simulation was performed with a Planck 2018~\cite{Planck2018} flat $\Lambda$CDM cosmology noted \texttt{cosmo000}. The redshift of the simulation used is $z=0.2$ because the Halo Occupation Distribution (HOD) model used to simulate BGS was tuned at $z=0.2$ and \texttt{AbacusSummit} was not run until $z=0.0$. This means that the underlying density field is less clustered than at the present time, and needs to be accounted for in the fitting procedure. The true growth rate associated with this simulation is derived from the Boltzmann solver \texttt{CLASS} \git{lesgourg/class_public}{}~\citep{class2011} and is $\fsifid = 0.4622$ at the redshift of the simulation ($z=0.2$).

The main simulation box size is sufficient to contain 27 times the overlapping volume between DESI BGS and ZTF \sn~fields. As pointed out in \textbf{C23} and in~\cite{Amenouche2024}, the ZTF survey is volume-limited for redshifts $z \lesssim 0.06$. Consequently, without introducing Malmquist bias correction, it is necessary to use \sne~with redshift $z\lesssim 0.06$ to measure $\fsi$~with ZTF. The subdivision of the large \texttt{AbacusSummit} into 27 sub-boxes provides simulations with objects located up to $z_{\mathrm{max}} = 0.11$, which is sufficient for the present study.

\begin{figure*}
	\includegraphics[width=\textwidth]{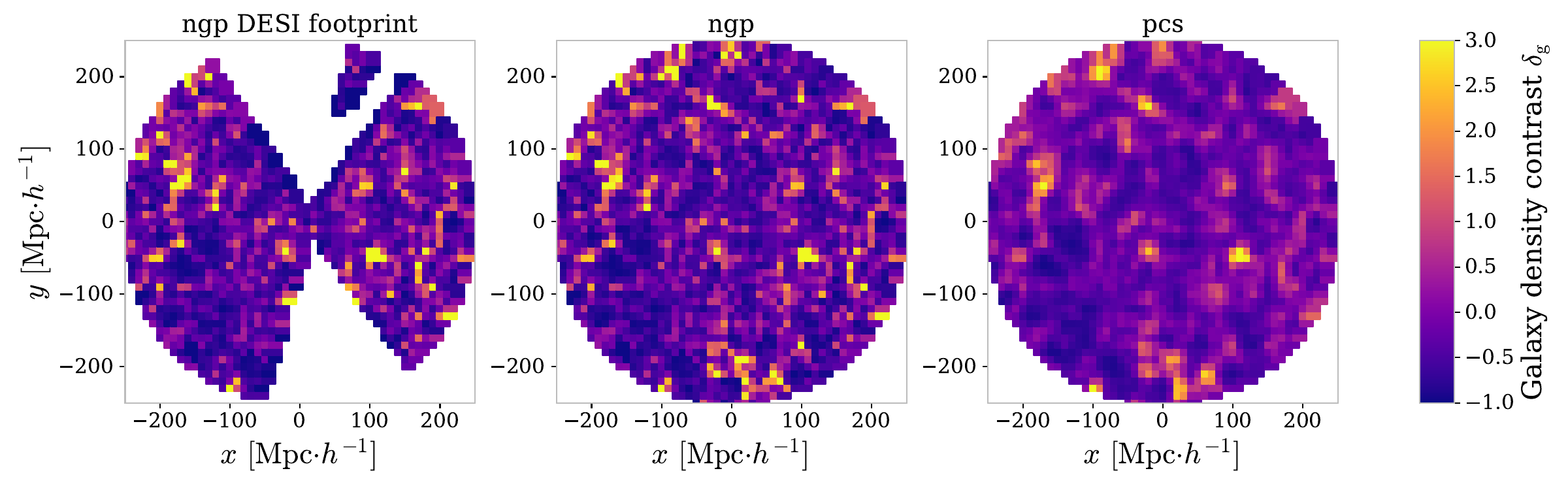}
    \caption{Slice through a spherical DESI BGS density mesh of radius 250~$\mpc$, with a voxel size of 10~$\mpc$. (left) Nearest grid point (NGP) interpolation scheme taking into account the DESI BGS footprint. (middle) NGP interpolation scheme without footprint. (right)  piecewise cubic spline (PCS) interpolation scheme without footprint.}
    \label{fig:meshes}
\end{figure*}

The generation of the DESI BGS mock catalog is given in detail in~\cite{Bautista2025}, but we provide a concise summary of the key ingredients. The \texttt{AbacusSummit} halo catalog is populated with a luminosity-dependent halo occupation distribution following~\cite{Smith2017,Smith2022}. To overcome the lack of low-mass halos in the \texttt{AbacusSummit} simulation, a random selection of field particles not associated with halos is performed. This procedure was tested in~\cite{Smith2023} and was shown to be a good approximation on particle cross-correlation statistics with halos. The number density of BGS is generated to match the projected correlation function of the DESI Data Release 1 BGS sample~\cite{DESI2025c}, extrapolated to the Data Release 3 (DR3) of DESI BGS given from target selection and magnitude cuts~\cite{Hahn2022}. The DR3 sample was chosen as it is the end of the initially planned DESI survey\footnote{DESI will now be a 8-year survey covering 17,000 $\mathrm{deg}^2$} and will contain a uniform volumetric density, avoiding potential biases caused by regions poorly observed. Random catalogs are generated from a uniform sampling over the whole volume. An assignment scheme with existing BGS galaxies is used to ensure that the distribution of the random galaxies follows that of BGS. The resulting BGS catalog is shown in figure~\ref{fig:density_field} for one mock.

As the total number of BGS galaxies for each sub-box is on the order of 1.8 million with the DESI DR3 footprint and 5.2 million without it, it is not computationally possible to work at the field level with individual galaxies. We perform a meshing of the density field using the \texttt{pmesh}~\git{rainwoodman/pmesh}{} software. As this step can modify the clustering of galaxies, and thus the inferred growth rate, we test various mesh configurations:

\begin{itemize}
    \itemsep0em 
    \item \textbf{Mesh shape}: spherical or cubic cut in Cartesian coordinates
    \item \textbf{Total mesh size (noted R)} in terms of radius for spherical mesh or half side-length for cubic mesh: 150, 200, 250, or 300~$\mpc$
    \item \textbf{Mesh voxel size (noted L)}: 10, 20, or 30~$\mpc$
    \item \textbf{Mesh assignment scheme}: nearest grid point (NGP), cloud-in-cell (CIC), triangular-shaped cloud (TSC), or piecewise cubic spline (PCS)
    \item \textbf{Footprint}: considering the DESI BGS DR3 footprint or not
\end{itemize}

Figure~\ref{fig:meshes} shows an illustrative example of mesh configurations varying footprints and assignment schemes. The choice of mesh assignment changes the number of galaxies that are assigned to a given mesh cell, effectively changing the smoothing of the resulting density field. The different mesh parameter variations constitute a total of 192 different mesh configurations. We compute the galaxy density field in cell $i$ as

\begin{equation}
    \label{eq:density}
    \delta_{\mathrm{g},i} = \left( \frac{N_{\mathrm{cell},i} /  \sum_i N_{\mathrm{cell},i}}{N_{\mathrm{exp},i} /  \sum_i N_{\mathrm{exp},i}}\right)  - 1\, ,
\end{equation}

\noindent where $N_{\mathrm{cell},i}$ is the number of galaxies associated with a cell depending on the assignment scheme. $N_{\mathrm{exp},i}$ is the expected number of galaxies estimated from the number of galaxies in the random catalogs. The sums are run over all the mesh cells to normalize for different number of particles in the random and galaxy catalogs. The uncertainty associated with the density field is modeled via the Poisson statistics as 

\begin{equation}
    \label{eq:density_err}
    \sigma_{\delta,\mathrm{g},i} = \frac{1}{\sqrt{N_{\mathrm{exp},i}/ N_{\rm random}}}\, ,
\end{equation}

\noindent where $N_{\rm random}$ is the number of random catalogs used to compute the window function of the survey.

\subsection{ZTF supernovae simulation}
\label{subsec:ztfsim}

We simulated ZTF \sne~using the \texttt{snsim}$^{\ref{bastiencarreres/snsim}}$ library following \textbf{C23}. These simulations aim to reproduce the spectroscopically-classified SN Ia sample of ZTF at the end of the 6-years survey. The number of events is drawn from the Poisson law with parameter

\begin{equation}
    \lambda = R_\text{SN}  \times T \times V,
\end{equation}

\noindent where $T$ is the duration of the survey and $V$, its volume. $R_\text{SN}$ is the SN~Ia rate in units of time and volume. We used the value of \cite{perley_zwicky_2020} re-scaled to our value of $h$

\begin{equation}
    R_\text{SN} =  2.35 \times 10^{-5} \left(\frac{h}{0.70}\right)^3 \text{Mpc}^{-3} \text{yr}^{-1}\, .
\end{equation}

The time dilation is taken into account by integrating the rate corrected by a factor of $(1+z)^{-1}$ over the volume of the survey. The Abacus DESI BGS catalog contains quality cuts from the DESI BGS survey. The \sn~host catalog is drawn from these BGS mocks. As the cuts and selections in the BGS catalogs could potential affect the rate of \sn~explosion, we enforce a volumetric rate for \sne~by assigning a weighted probability to each galaxy. The \sne~luminosities are modeled using the SALT2 spectral template \citep{guy_salt2_2007}, which is parametrized by the stretch, $x_1$, the color, $c$, and the magnitude in the rest-frame Bessell-B band, $m_B$. The $x_1$ values are drawn from the distribution described in \cite{nicolas_redshift_2021}, and the $c$ values are drawn from the asymmetric Gaussian described in Table~1 \cite{scolnic_measuring_2016} (low-$z$). We fixed the absolute magnitude of \sne~to the value found in \cite{betoule_improved_2014}, rescaled to our fiducial $h$, as

\begin{equation}
    M_B = -19.05 + 5 \log\left(\frac{h}{0.7}\right)\, .
\end{equation}

Using the Tripp relation \citep{Tripp1998}, the apparent peak-magnitude of each \sn~is given by 

\begin{equation}
    m_{B, i} = M_B - \alpha x_{1,i} + \beta c_i + \mu(z_{\text{cos},i}) + 10\log(1 + z_{p,i})\, ,
\end{equation} 

\noindent where $\mu(z_{\text{cos},i})$ is the distance modulus evaluated at the cosmological redshift and the last term is the relativistic beaming due to PV. We note that this Tripp relation does not contain a broken-$\alpha$ stretch distribution~\cite{Ginolin2025} or an environmental tracer term for the \sn~magnitude (such as a mass step). We choose to leave these considerations for future data-focused analyses. 

The time of peak magnitude $t_0$ of the \sne~in Bessell-B band is uniformly generated over the survey duration. To simulate realistic observations, the explosion time and angular position distribution of \sne~are matched to the existing observations of ZTF. Fluxes are then generated using the \texttt{sncosmo} \git{sncosmo/sncosmo}{} python package within the corresponding observed band and observational noise. The error model for fluxes is

\begin{equation}
    \sigma_F^2 = \frac{F}{g} + \sigma_\text{sky}^2 + \left(\frac{\ln10}{5}F\right)^2 \sigma_\text{zp}^2 \, ,
\end{equation}

\noindent where the first term is the Poisson error, with $g$ the gain in photo-$e^-$ / ADU. The parameter $\sigma_\text{sky}$ is the sky noise of the observation and  $\sigma_\text{zp}$ is the zeropoint error fixed at $\sigma_\text{zp} = 0.01$~mag to match the ZTF data error bars~\cite{carreres_growth-rate_2023, Amenouche2024}.
With this procedure, we simulate ZTF-like lightcurves for the equivalent of 6 years of the survey.

After the simulation, we applied the selection of the ZTF spectroscopic follow-up as in \textbf{C23} in order to reproduce the ZTF selection function described in \cite{Bellm2019} and \cite{Rigault2024}. We fitted the generated light curves with the SALT2 spectral template to obtain the estimated parameters $m_B$, $x_1$, and $c$, along with their associated covariance matrix,

\begin{equation}
    C_\text{SALT} = \begin{pmatrix}
        \sigma_{m_B}^2 & \text{Cov}[m_B,x_1] & \text{Cov}[m_B,c] \\
        \text{Cov}[m_B,x_1] & \sigma_{x_1}^2 & \text{Cov}[x_1,c] \\
        \text{Cov}[m_B,c]  & \text{Cov}[x_1,c] & \sigma_{c}^2 
    \end{pmatrix}.
\end{equation}

We applied quality cuts to SALT2 fit outputs. Following \textbf{C23}, we require that the light curve has at least 3 epochs in an interval of ten days around maximum brightness and a probability of the SALT fit greater than 95 \%. We also cut errors on $x_1$ and $t_0$ such that $\sigma_{x_1} < 1$ and $\sigma_{t_0} < 1$. Finally, we apply stretch/color cuts: $|x_1| < 3$ and $|c| < 0.3$.

We build the \sn~Hubble diagram by estimating the distance modulus as
\begin{equation}
     \mu_{\text{obs},i} = m_{B,i} + \alpha x_{1,i} - \beta c_i - M_0\, ,
\end{equation}

\noindent where $\alpha$, $\beta$ and $M_0$ are free parameters that will be fitted and marginalized jointly with the growth rate inference (see section~\ref{sec:results}). The velocities are estimated from the Hubble diagram residuals defined as
\begin{equation}
     \label{eq:hdresiduals}
     \Delta \mu_i(\alpha, \beta, M_0) = \mu_{\mathrm{obs},i}(\alpha, \beta, M_0) - \mu_{\mathrm{model},i}(z_{\text{obs}, i})\, .
\end{equation}

The errors associated with these residuals are computed from SALT covariance matrix as 

\begin{equation}
    \label{eq:velerror}
    \sigma_{\Delta \mu,i}^2(\alpha, \beta, \sigma_M) = \irow{1&\alpha&-\beta} \cdot C_{\mathrm{SALT},i} \cdot \irow{1&\alpha&-\beta}^T +\sigma_M^2\, ,
\end{equation}

\noindent where $\sigma_M$ is the \sn~intrinsic scatter which is treated as a free parameter in our model and assumed to be color-independent. Finally, the peculiar velocities are determined following the estimator derived in \textbf{C23}:

\begin{equation}
    \label{eq:velestimator}
    v_i = -\frac{c \ln(10)}{5} \left(\frac{(1+z_{\mathrm{obs},i})c}{(H(z_{\mathrm{obs},i})/h) D(z_{\mathrm{obs},i})} - 1 \right)^{-1} \Delta\mu_i\, ,
\end{equation}

\noindent with associated error bars $\sigma_{v_i}$ computed by replacing $\Delta\mu_i$ by $\sigma_{\Delta \mu,i}$ in equation \ref{eq:velestimator}. A fiducial cosmological model is used to compute this peculiar velocity estimator. We leave the study of parameter dependency of this estimator for future work. We assume that there is no error on \sn~redshifts, which will typically be the case with dedicated host redshift programs (see e.g.,~\cite{Soumagnac2024}) for which redshift errors are negligible compared to distance estimator errors.

The generated \sn~sample follows the same distribution as figures 3 and 4 of \textbf{C23}~\cite{carreres_growth-rate_2023}. For all the \sn~simulation performed on the 27 \texttt{AbacusSummit} halo catalogs, we verified that the total number of \sne~and their distribution (angular and radial) were the same as \textbf{C23}. We choose not to reproduce the figures showing this distribution to avoid duplication.

\section{Likelihood-based field-level estimator}
\label{subsec:method}

The fit of the growth rate parameter, $\fsi$, is performed using the likelihood-based field-level inference method implemented in the \texttt{flip} software and validated in \cite{Ravoux2025}. This software generates theoretical field covariance  in an algorithmically-optimized way of the general form

\begin{equation}
C_\mathrm{ab}(\mathbf{r_a},\mathbf{r_b}) = \frac{1}{(2 \pi)^3} \int d^3 \mathbf{k}  P_{\mathrm{ab}}(k, \mu_a, \mu_b) e^{i \mathbf{k}\cdot\mathbf{r}}\, ,
\end{equation}

\noindent between two fields $a$ and $b$ for any power spectrum model which can be expressed as 

\begin{equation}
\label{eq:Pwflip}
P_{\mathrm{ab}}(k, \mu_a, \mu_b) = \sum_{n} w_{\mathrm{ab},n} F_{\mathrm{ab},n}(k, \mu_a, \mu_b)  \mathcal{P}_{\mathrm{ab},n}(k) \, ,
\end{equation}

\noindent where $w_{\mathrm{ab},n}$ are model parameters to minimize, $F_{\mathrm{ab},n}$ are geometrical terms integrated analytically, $\mathcal{P}_{\mathrm{ab},n}$ are power spectrum terms integrated numerically, and $\mu_i = \mathbf{k}\cdot\mathbf{r}_i / (kr_i)$. The latter can carry additional parameters when covariance interpolation is used. Because our data span a wide sky area at low redshift, we include wide-angle effects in the power spectrum model. When wide-angle effects are considered in the model, the covariance matrix is computed as a linear decomposition of Hankel transforms ($\mathcal{H}$):

\begin{equation}
      \label{eq:cov_flip_wa}
       C_\mathrm{ab}(\mathbf{r_a},\mathbf{r_b})  = \sum_n w_{\mathrm{ab},n}  \sum_{\ell, \ell_1, \ell_2}N_{\mathrm{ab},\ell}^{\ell_1,\ell_2}(\phi,\alpha) \mathcal{H}_\ell\left[\mathcal{P}_{\mathrm{ab},n}(k) M_{\mathrm{ab},n}^{\ell_1, \ell_2}(k)\right](r)\, ,
\end{equation}

\noindent where $N_{\mathrm{ab},\ell}^{\ell_1,\ell_2}$ is a purely geometrical term accounting for wide-angle effects, and $M_{\mathrm{ab},n}$ is an analytical integration of the $F_{\mathrm{ab},n}(k, \mu_a, \mu_b)$ terms. We refer the reader to~\cite{Ravoux2025} for the mathematical definition of those terms and for more details on this calculation.

We use the \textbf{RC25} model from~\cite{Ravoux2025}, which includes robust treatment of wide-angle effects and non-linear small-scale clustering. This model is specialized to the fields considered (density, velocity, or combined fit). The full power spectrum model is given by:

\begin{equation}
\label{eq:model}
\begin{aligned}
& \mathcal{P}_{\mathrm{gg}}=\left[(\bsi)^2 P_{\mathrm{mm}}(k)+(\bsi) (\fsi) \left(\mu_a^2+\mu_b^2\right) P_{\mathrm{m} \theta}(k)+(\fsi)^2 \mu_a^2 \mu_b^2 P_{\theta \theta}(k)\right] \exp \left[\frac{-k^2\left(\mu_a^2+\mu_b^2\right) \sigg^2}{2}\right] \, , \\
& \mathcal{P}_{\mathrm{gv}}=i a H \frac{\mu_b}{k}\left[(\bsi) (\fsi) P_{\mathrm{m} \theta}(k)+(\fsi)^2 \mu_a^2 P_{\theta \theta}(k)\right] \exp \left[\frac{-k^2 \mu_a^2 \sigg^2}{2}\right] D_{\mathrm{u}}\left(k, \sigu\right) \, ,\\
& \mathcal{P}_{\mathrm{vv}}=(aH)^2 (\fsi)^2 \frac{\mu_a \mu_b}{k^2} P_{\theta \theta}(k) D_{\mathrm{u}}^2\left(k, \sigu\right) \, ,
\end{aligned}
\end{equation}

\noindent where $b$ is the galaxy bias, $\sigg$ is parameter used in the Gaussian model of RSD Finger-of-God damping on galaxy clustering, caused by randomized velocities at small scales. $D_{\mathrm{u}}$ is a damping function used in peculiar velocity studies to model the impact of RSD Finger-of-God on the positions of peculiar velocities (see e.g., \cite{koda_are_2014}) parameterized by $\sigu$. Power spectra $P_{\mathrm{mm}}$, $P_{\mathrm{m\theta}}$ and $P_{\mathrm{\theta\theta}}$ are associated to the matter $m$ and velocity divergence $\theta$. Those power spectra are represented in figure~\ref{fig:power_spectra} and are computed following~\cite{bel_accurate_2019}:

\begin{figure}
	\includegraphics[width=\columnwidth]{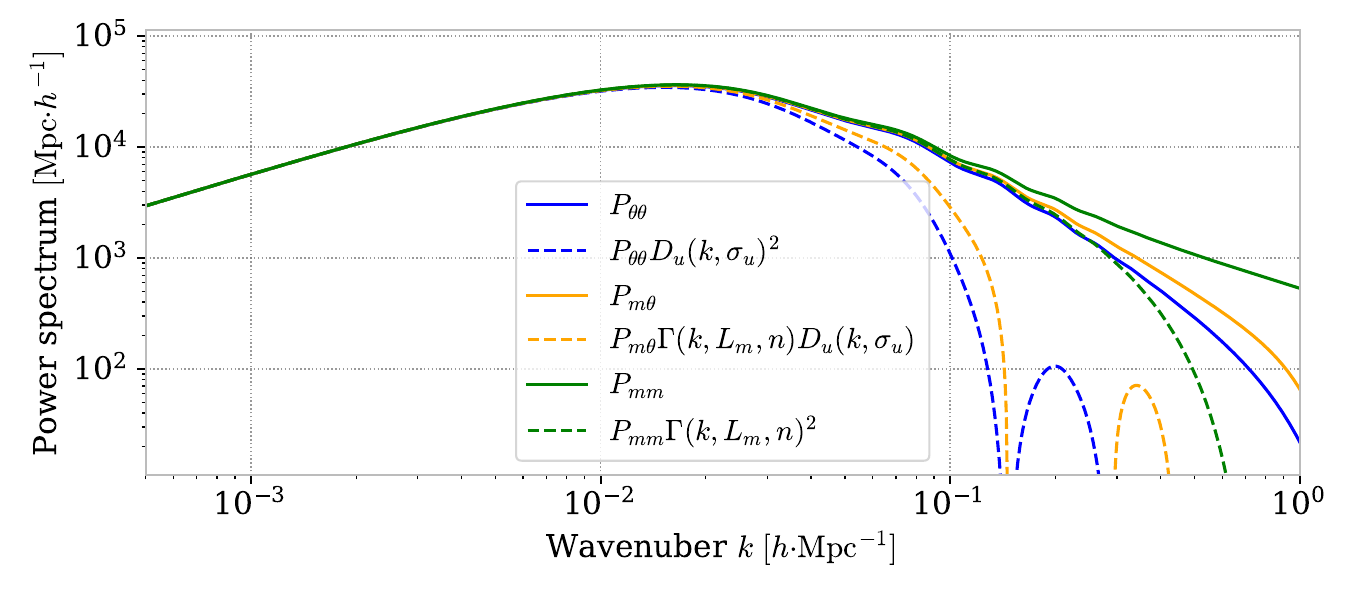}
    \caption{Power spectra used in the \texttt{flip} package to compute $gg$, $gv$ and $vv$ field-covariance matrices.  All input power spectra are obtained from the Boltzmann solver \texttt{CLASS} with the~\cite{bel_accurate_2019} parameterization (detailed in equation \ref{eq:bel}). We assume the Planck 2018~\cite{Planck2018} flat $\Lambda$CDM cosmology \texttt{cosmo000} used in \texttt{AbacusSummit} simulations for the linear power spectrum calculation. Plain lines show the input velocity divergence power spectrum, $P_{\theta\theta}$ (blue), the matter power spectrum, $P_{\rm mm}$ (green), and the cross-power spectrum, $P_{\rm m \theta}$ (yellow). The dashed lines show the same power spectra after adding the damping terms which accounts for meshing sampling ($\Gamma(k, L, n)$ in equation \ref{eq:gridding_cor}, with an NGP mesh assignment and a $10\ \mpc$ mesh voxel size) and random displacement of velocity positions at small scales ($D_{u} (k, \sigu)$, taking $\sigu = 21.5\ \mpc$). The impact of mesh sampling and velocity dispersion is different on each power spectra, as per the figure legend and equation \ref{eq:model}. The impact of the FoG parameter $\sigg$ is not represented here as it has an anisotropic impact.}
    \label{fig:power_spectra}
\end{figure}

\begin{equation}
\label{eq:bel}
\begin{aligned}
& P_{\mathrm{mm}}(k)= \frac{P_{\mathrm{HF}}(k)}{\si^2} \, , \\
& P_{\mathrm{m\theta}}(k)=\frac{\left(P_{\mathrm{HF}}(k) P_{\mathrm{L}}(k)\right)^{\frac{1}{2}}}{\si^2} \exp\left[-\frac{k}{k_\delta (\si)}-b(\si) k^6\right] \, ,\\
& P_{\mathrm{\theta\theta}}(k)= \frac{P_{\mathrm{L}}(k)}{\si^2} \exp\left[-k\left(a_1(\si)+a_2(\si) k+ a_3(\si) k^2\right)\right] \, ,
\end{aligned}
\end{equation}

\noindent where the $k_\delta$, $b$, $a_1$, $a_2$, and $a_3$ are the fitted functions of~\cite{bel_accurate_2019} depending on $\si$, $P_{\mathrm{HF}}$, and $P_{\mathrm{L}}$, which are respectively the non-linear halofit and linear matter power spectra computed using the Boltzmann solver \texttt{CLASS} and the cosmological model of the simulation (\texttt{cosmo000}). 

In our model, the parameters of interest ($f$ and $b$) are totally degenerate with the amplitude of the linear power spectrum. Consequently, all power spectra are renormalized by the fiducial $\si$ value, and the fitted parameters are the combinations $\fsi$ and $\bsi$. For all calculations, the $aH$ term is computed for the redshift of the simulation output, i.e. $z=0.2$. For consistency, the measured $\fsi$ value is compared to its fiducial value $\fsifid$ at the same redshift.

We do not consider potential variations or biases on the cosmological parameters used in the determination of distances and matter power spectrum calculations. The $\si$ normalization chosen in the field covariance model tends to minimize the impact of a mismatch for a parameter which impacts the amplitude of the input power spectrum. However, other finer effects such as wavenumber-dependent parameter variation or the impact on the distance calculation is not considered here. This study is out of scope of this paper and will be performed in future studies, in particular, for the impact of $H_0$, $\Omega_m$, and $n_{\mathrm{s}}$ fiducial values.

The fields considered in this study are the peculiar velocity, $v$, and galaxy density, $\delta_{\mathrm{g}}$. The galaxy density is derived from equation~\ref{eq:density} with different configurations but without any explicit free parameter. The meshing of the density field induces a smoothing of the correlation at small scales depending on the mesh geometry. This smoothing is introduced in our model by multiplying $P_{\mathrm{m\theta}}$ by $\Gamma$ and $P_{\mathrm{mm}}$ by $\Gamma^2$, where the mesh window function is defined by:

\begin{equation}
\label{eq:gridding_cor}
\Gamma(k, L, n) = \frac{1}{4\pi} \left(\frac{8}{L^{3}}\right)^n \int \left[\frac{\sin\left( \frac{k_x L}{2} \right)}{k_x}
\frac{\sin\left( \frac{k_y L}{2} \right)}{k_y}
\frac{\sin\left( \frac{k_z L}{2} \right)}{k_z}\right]^n  \sin\theta d\theta d\phi\, ,
\end{equation}

\noindent where $L$ is the mesh voxel size, $n$ is the order associated with the mesh assignment scheme (1 for NGP, 2 for CIC, 3 for TSC, and 4 for PCS), $k_x = k \sin\theta \cos\phi$, $k_y = k \sin \theta \sin \phi$, and $k_z = k\cos\theta$.

The figure~\ref{fig:power_spectra} shows the impact of the mesh window function and the $D_{\mathrm{u}}$ damping function on the different power spectrum terms, for $\sigu = 21.5\ \mpc$, an NGP mesh assignment, and a $10\ \mpc$ mesh voxel size for densities. 

For the velocity field, we consider two variations either with the true velocity field from the simulation in a non-realistic situation, or derived from the results of \sn~light curve fitting. In the latter, the velocity field is derived from equation~\ref{eq:velestimator} and depends on the standardization parameters $\left\{\alpha, \beta, M_0 \right\}$. Since the number of \sne~used is relatively low (of the order of $\sim 1,700$), the velocity field is not sampled on a mesh and no window correction is applied.

Following \textbf{C23}, we add a simple observed covariance matrix to the velocity to account for the random dispersion of velocities and associated uncertainties:

\begin{equation}
    C_{vv,\mathrm{tot}} = C_{vv} +\left[\sigma_v^2+\sigma_{v,i}^2(\alpha,\beta,\sigma_M)\right] I_D\,,
\end{equation}

\noindent where $I_D$ is the identity matrix, $\sigma_v$ is a free parameter accounting for velocity dispersion, and $\sigma_{v,i}$ is the velocity error. The latter is zero when true velocities are considered. When realistic \sn~velocities are considered, $\sigma_{v,i}$ is given by replacing $\Delta\mu_i$ in equation~\ref{eq:velestimator} with $\sigma_{\Delta \mu,i}$ from equation~\ref{eq:velerror}. In that case, $\sigma_M$ is added to the list of parameters to fit.

For densities, the shot noise uncertainty estimated with equation~\ref{eq:density_err} is added to the density covariance matrix as

\begin{equation}
    C_{gg,\mathrm{tot}} = C_{gg} +\sigma_{\delta,\mathrm{g},i}^2 I_D\,.
\end{equation}

We use the multivariate Gaussian likelihood:

\begin{equation}
    \mathcal{L}[x
    (\Theta), C_{xx,\mathrm{tot}}(\Theta)] = (2\pi)^{-N_x/2}|C_{xx,\mathrm{tot}}(\Theta)|^{-\frac{1}{2}} \exp\left[-\frac{1}{2}x(\Theta)^TC_{xx,\mathrm{tot}}(\Theta)^{-1}x(\Theta)\right]\, ,
\end{equation}

\noindent where $\Theta$ are the parameters to fit (both cosmological and nuisance) and the shape of the covariance matrix $C_{xx,\mathrm{tot}}$ depends on the considered field $x$, of size $N_x$, which can be $\delta$, $v$, or both $x = \irow{\delta& v}$. For the combined case, the covariance matrix is defined as:

\begin{equation}
    C_{xx,\mathrm{tot}} = \begin{bmatrix}
        C_{\delta\delta,\mathrm{tot}} & C_{\delta v} \\
        C_{v \delta}     & C_{vv,\mathrm{tot}} 
    \end{bmatrix}\, .
\end{equation}

In addition, the covariance matrix can be interpolated for parameters which are not present as coefficients of the linear decomposition of the power spectrum model in equation~\ref{eq:Pwflip}. This linear interpolation allows for fitting one additional parameter beyond $w_{\mathrm{ab},n}$, but slows down the fitting procedure as it manipulates a large number of covariance matrices. For the density fit, the covariance calculation is interpolated for the $\sigg$ parameter. For velocity and combined fits, we interpolate the $\sigu$ parameter, which is the main contaminant of $\fsi$. We chose to not fit for the $\sigg$ parameter and to fix it from the results of the density fit only. Including this parameter in the combined fit requires a 2D interpolation of a very large covariance matrix and is numerically out of reach with the current likelihood configuration; we will let this configuration for a future study. As a summary, we provide the list of parameters fitted (cosmological, nuisance, and astrophysical) depending on the type of fit performed in this paper:

\begin{itemize}
    \itemsep0em 
    \item \textbf{True velocity fit}: $\Theta = \left\{\fsi, \sigma_{v}, \sigu\right\}$
    \item \textbf{\sn~velocity fit}: $\Theta = \left\{\fsi, \sigma_{v}, \sigu,\alpha,\beta,M_0,\sigma_M \right\} $
    \item \textbf{Density fit}: $\Theta = \left\{\fsi, \bsi, \sigg \right\} $
    \item \textbf{Combined fit}: $\Theta = \left\{\fsi,\bsi, \sigma_{v}, \sigu,\alpha,\beta,M_0,\sigma_M \right\} $

\end{itemize}

For all fit, the chosen likelihood is either minimized with a dedicated \texttt{iminuit} \git{scikit-hep/iminuit}{}~\citep{minuit1975,iminuit2020} fitter integrated inside \texttt{flip}, or sampled with a Markov chain Monte Carlo (MCMC) method with the \texttt{emcee} \git{dfm/emcee}{}~\citep{emcee} software. The cost of MCMC sampling being very large, we make the \texttt{iminuit} fit for all the configurations we explore, and we reserve the MCMC fit for the chosen configuration for which we seek to characterize correlations between parameters. For all the fits presenter hereafter, either with minimization or sampling, the limits and priors adopted are given in table~\ref{tab:priors}.

\begin{table}[htbp]
\centering
\begin{tabular}{@{}llcc@{}}
\toprule
\textbf{Parameter} & \textbf{\texttt{iminuit} limits} & \textbf{\texttt{iminuit} priors} & \textbf{MCMC priors} \\
\midrule
$\fsi$              & [0, +$\infty$] & -- & $\mathcal{U}\left(0.1, 2 \right)$\\
$b\sigma_8$         & [0, +$\infty$] & -- & $\mathcal{U}\left(0.1, 2\right)$\\
$\sigu$ no prior & [2, 50] & -- & no MCMC fit\\
$\sigu$          & [11, 32] & $\mathcal{N}\left(21.5,10^2\right)$ &  $\mathcal{N}\left(21.5,10^2\right)$\\
$\sigg$          & [1, 25] & -- & $\mathcal{U}\left(1, 25 \right)$\\
$\sigma_v$          & [0, 3000] & -- & $\mathcal{U}\left(0, 500 \right)$\\
$\alpha$            & [0.1, 0.2] & -- & $\mathcal{U}\left(0.05, 0.25 \right)$\\
$\beta$             & [2.5, 4.5] & -- & $\mathcal{U}\left(2, 4\right)$\\
$M_0$               & [-20, -18] & -- & $\mathcal{U}\left(-20, -18 \right)$\\
$\sigma_M$          & [0.05, 0.2] & -- & $\mathcal{U}\left(0.08, 0.2 \right)$\\
\bottomrule
\end{tabular}
\caption{Limits on the \texttt{iminuit} fits and priors used both for \texttt{iminuit} and MCMC fits when the parameter is varied. For $\sigu$, two cases are considered if the fit is a blind search of the true $\sigu$ value on the simulation or for standard fits. Gaussian priors are noted as
$\mathcal{N}(\mu,\sigma^2)$.}
\label{tab:priors}
\end{table}

\section{Growth rate fit results}
\label{sec:results}

\subsection{BGS density field fit}
\label{subsec:density}

We perform field-level fits from the density field of the simulated DESI BGS DR3 galaxy catalog. This step allows us to test our inference only on the matter field, and to explore the space of possible density field configurations to characterize their biases and error bars. We perform a best-fit \texttt{iminuit} minimization with an interpolation of the covariance matrix over the $\sigg$ parameter. Those fits are done over the 27 DESI BGS mocks, for all the 192 mesh configurations detailed in the previous section, as well as two maximal integration wavenumbers $\kmax = 0.2$ or $1.0$~\hpmpc.

\begin{figure}
    \centering
	\includegraphics[width=0.49\columnwidth]{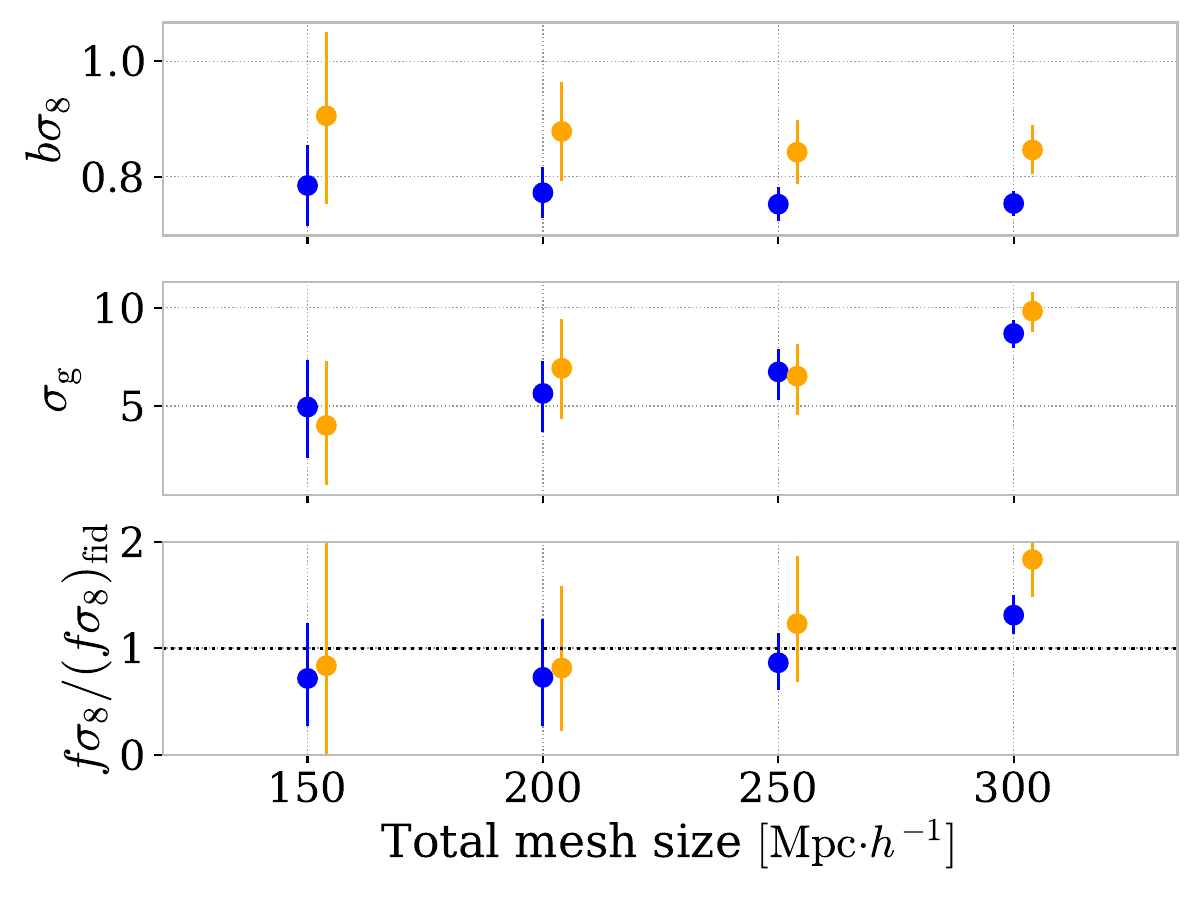}
	\includegraphics[width=0.49\columnwidth]{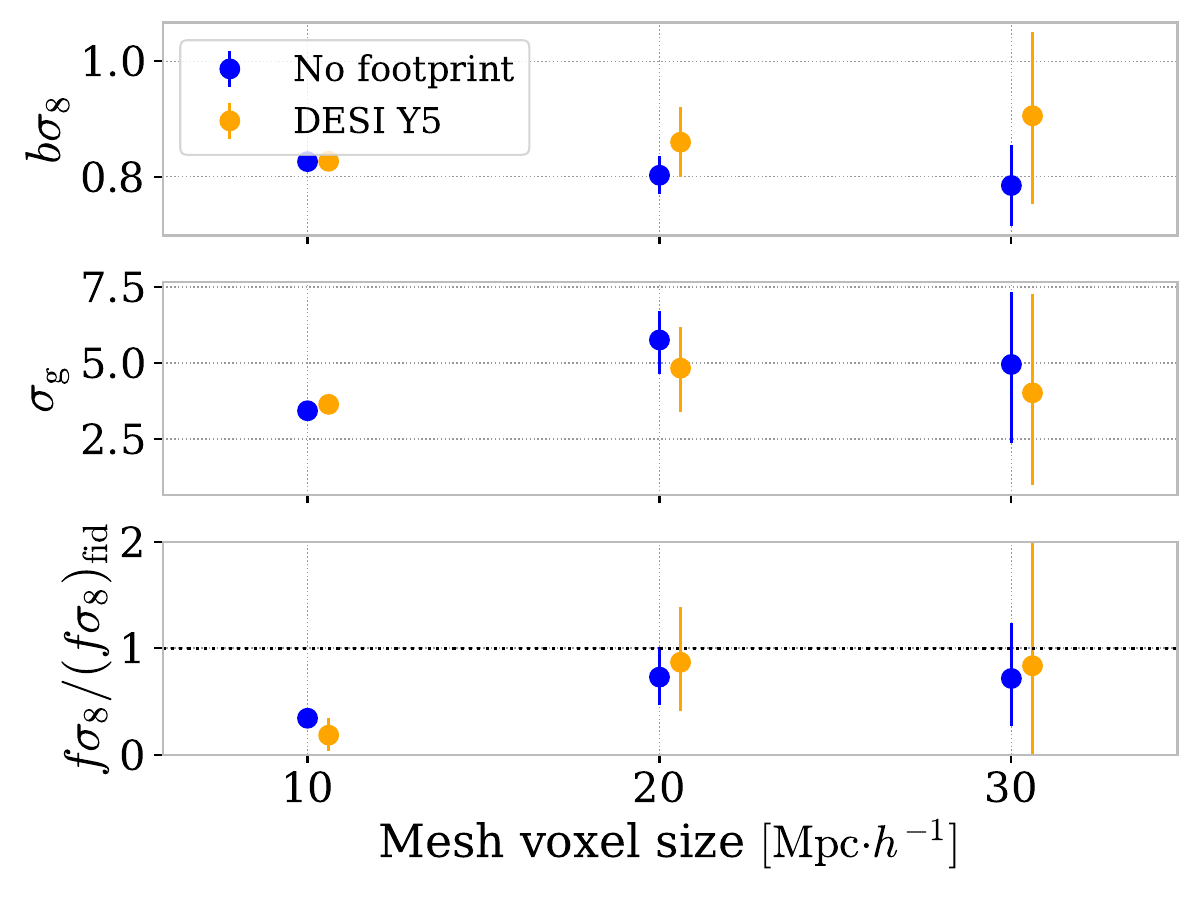}
    \caption{Field-level likelihood-based fit of the DESI BGS density field, meshed with different configurations. Starting from a baseline configuration ($\kmax = 1.0~\mpc$, 150 $\mpc$ total mesh size, 30 $\mpc$ mesh voxel size, spherical shape, and NGP assignment scheme), we vary one parameter at a time. The fitted values and errors shown are the average over the 27 BGS mocks. The fit is shown without footprint cut (blue) and with the DESI BGS footprint (orange). The dotted line shows the true expected value. (left) Variation of total mesh size. (right) Variation of mesh voxel size.}
    \label{fig:density_rcom_gridsize}
\end{figure}

\begin{figure}
    \centering
	\includegraphics[width=0.49\columnwidth]{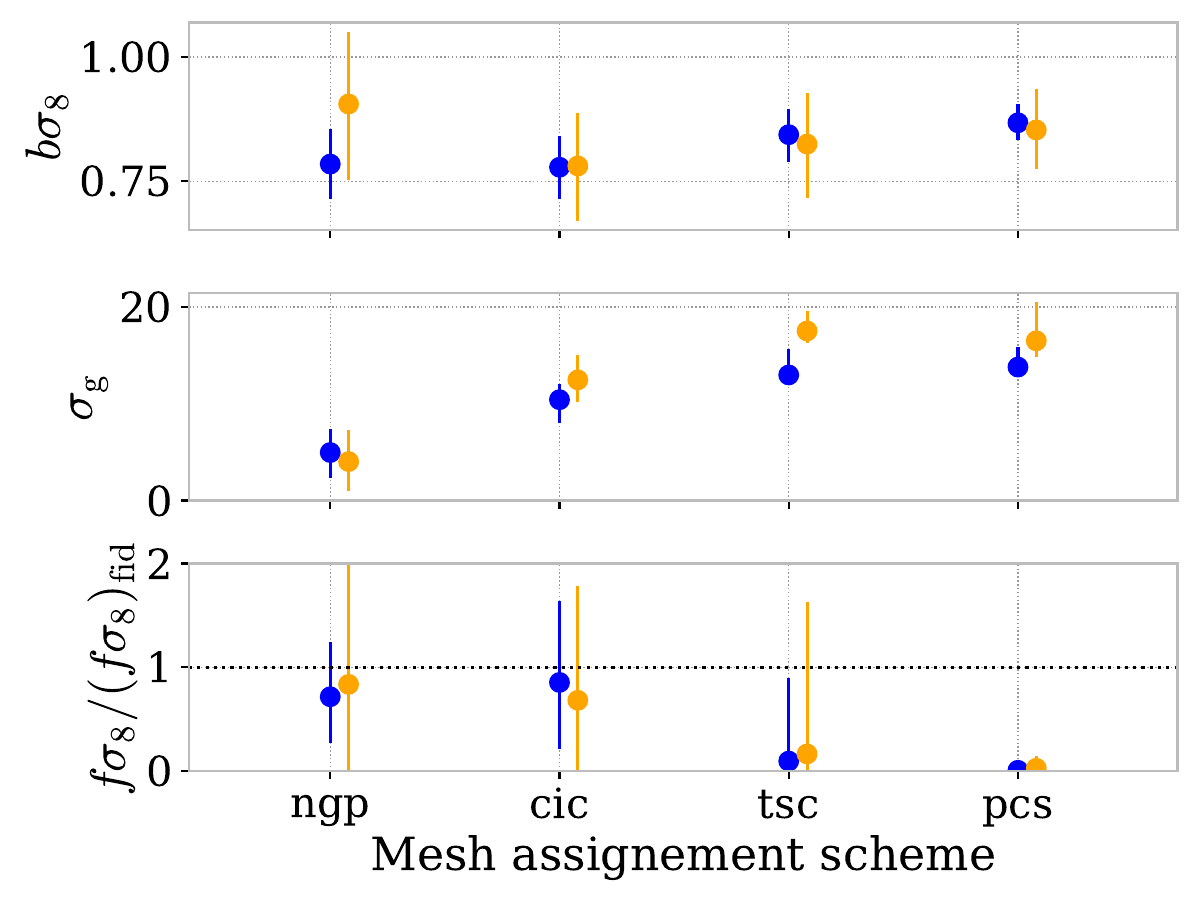}
	\includegraphics[width=0.49\columnwidth]{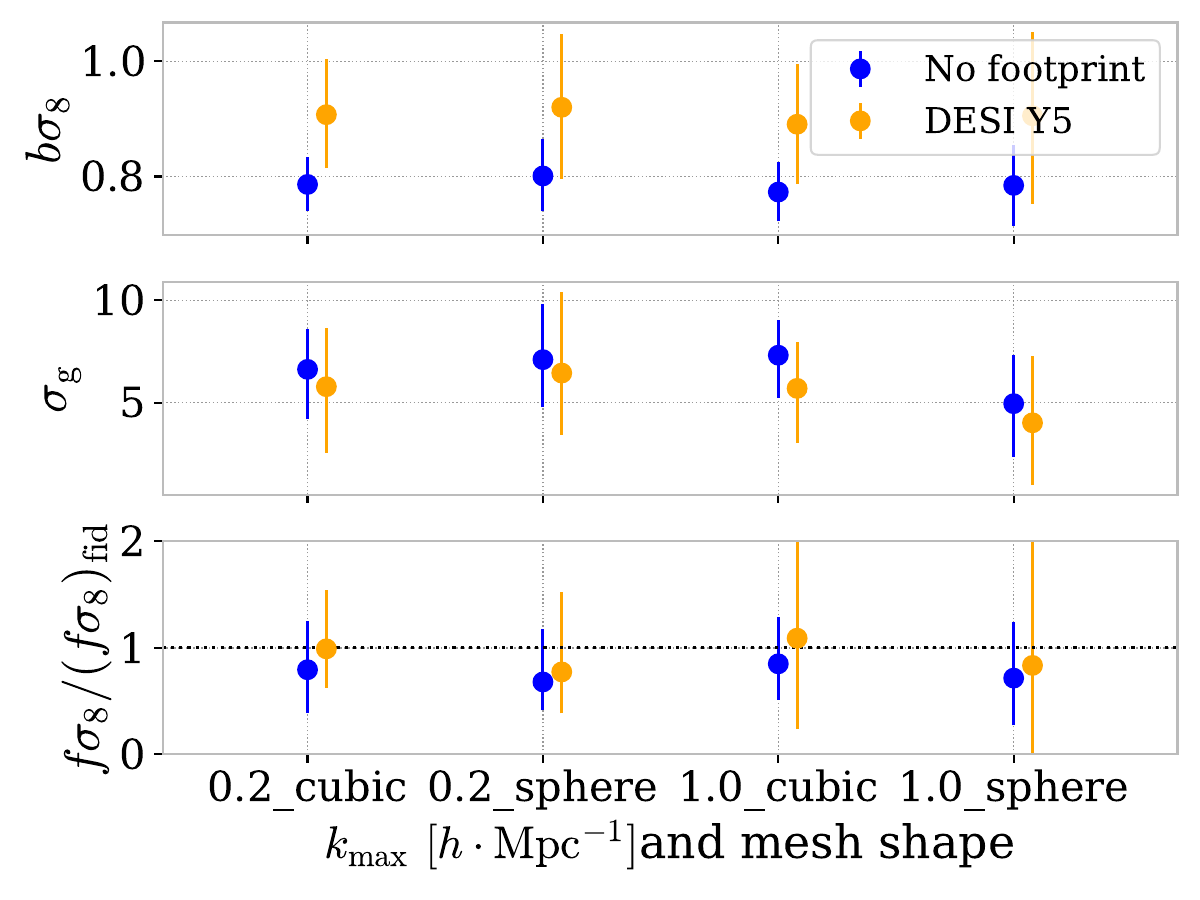}
    \caption{Same as figure~\ref{fig:density_rcom_gridsize} but varying different configuration parameters. (left) Mesh assignment scheme. (right) Maximal wavenumber $\kmax$ used in the field-covariance integration and shape of the mesh (spherical or cubic).}
    \label{fig:density_kind_kmax_gridtype}
\end{figure}

\begin{figure}
    \centering
	\includegraphics[width=0.7\linewidth]{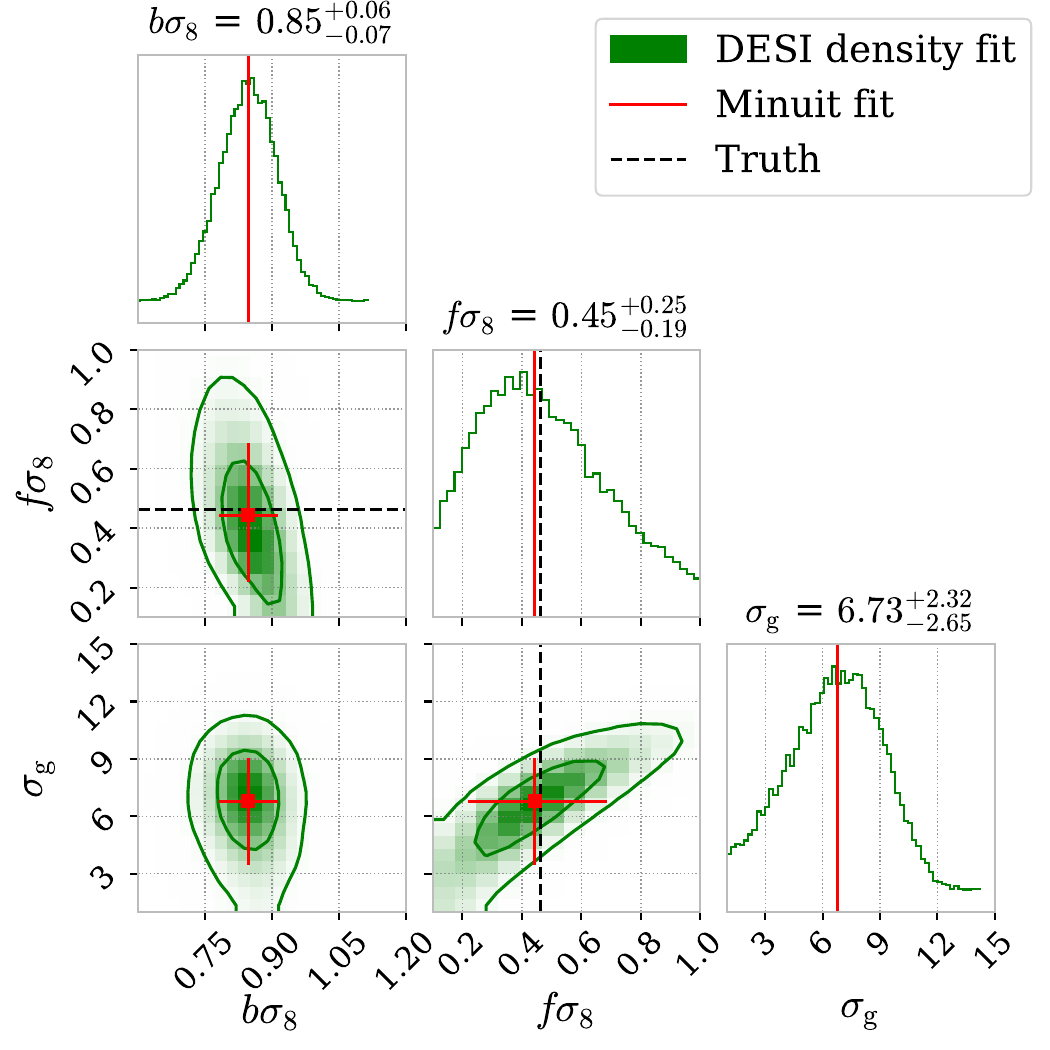}
    \caption{Markov Chain Monte Carlo sampling of the likelihood considering only the DESI BGS density field with footprint cut, for one simulation with the baseline mesh configuration (voxel size of $30~\mpc$, a spherical shape, an NGP assignment scheme, and a total size of $150~\mpc$), and $\kmax = 1.0$~\hpmpc. The red points and error bars show the \texttt{iminuit} best fit and \texttt{MINOS} error, and the true $\fsi$ value is shown with black dotted line.}
    \label{fig:density_mcmc}
\end{figure}

Figures~\ref{fig:density_rcom_gridsize} and~\ref{fig:density_kind_kmax_gridtype} show the results of the BGS density fits. We reject fits with invalid convergence criteria set by the \texttt{iminuit} software. To highlight the impact of each parameter, we start from a baseline configuration fit with a maximal integration wavenumber $\kmax = 1.0~\mpc$, a mesh with a voxel size of $30~\mpc$, a spherical shape, an NGP assignment scheme, and a total size of $150~\mpc$. We show the impact of varying only one configuration parameter from this baseline fit. Points on each figure show the weighted average of the 27 BGS mocks. The weights are the inverse variance of the fitted parameter computed with the \texttt{MIGRAD} routine of \texttt{iminuit}. The weighting avoids contamination from outliers or bad fits. The error bars are computed as the average per-mock \texttt{MINOS} error over the 27 mocks. In order to limit the number of very time-consuming fits, we only consider density fields with a number of elements lower than $15,000$. This cut excludes configurations with very large field elements, i.e., small voxel sizes and large total mesh sizes. All the tested configurations are shown in the appendix figures~\ref{fig:density_variation} and~\ref{fig:density_variation_2}. Finally, figure~\ref{fig:density_mcmc} shows the results of MCMC sampling for the baseline configuration of one specific mock. These fits were performed to obtain a clearer view on parameter correlations in the case of density field fits. 

Comparing the fit results in figures~\ref{fig:density_rcom_gridsize} and~\ref{fig:density_kind_kmax_gridtype} where DESI footprint is considered (orange points) or not (blue points), we conclude that the footprint cut mainly impacts the bias $b \sigma_8$ value while giving consistent values for $\fsi$ and $\sigg$. It indicates that our field-level method correctly accounts for the geometry of the survey without biasing $\fsi$. As expected, the footprint cut reduces the number of considered voxels and thus increases the mean error bar. 

Focusing on the impact of the total mesh size in figure~\ref{fig:density_rcom_gridsize} (left), for large meshes ($> 250\ \mpc$), the $\fsi$ value is biased. This bias is caused by the correlation between the $\fsi$ and $\sigg$ parameters. Indeed, the increase of $\fsi$ is concomitant with $\sigg$, and the result of the MCMC fit in figure~\ref{fig:density_mcmc} shows a clear correlation between both parameters. We verify that fixing the $\sigg$ value to the one fitted with a small total size suppresses the $\fsi$ bias. We interpret this change in $\sigg$ to be caused by the fact that the probed volume is sufficient to capture baryon acoustic oscillation (BAO) scales. In that case, the impact of non-linear smoothing of the BAO peak is also changing the $\sigg$ value. This interpretation agrees with the findings in~\cite{Ravoux2025}, which shows that for a galaxy fit on N-body simulations, the fitter adapts the $\sigg$ parameter to BAO peak smoothing, rather than the small-scale clustering. This conclusion shows a limitation of our linear model, which should be extended in future studies to have dedicated parameters for BAO peak smoothing and additional parameters for small-scale clustering. This extension, out of the scope of this paper, could be performed with EFTofLSS modeling (see e.g.,~\cite{Carrasco2013,Perko2016,DAmico2020,Ivanov2022}). For our case, we choose to fix the $\sigg$ value for coupled density-velocity fits presented later such that it does not introduce any bias in $\fsi$. 

Considering the effect of mesh voxel size in the right panel of figure~\ref{fig:density_rcom_gridsize}, the error bar on fitted parameters decreases as the voxel size decreases. However, the $\fsi$ value is clearly biased for $10~\mpc$ voxel size. This bias is expected as our linear model is inaccurate at such small scales. Indeed, for those scales, the impact of the small-scale power spectrum modeling is larger, and currently only $\sigg$ can constrain small-scale clustering, i.e., we are not covering the possible variations of the small-scale power spectrum. Similarly to the previous point, including non-linear theories such as EFTofLSS could solve this issue, potentially reducing the $\fsi$ error bars without introducing any bias.

The left panel of figure~\ref{fig:density_kind_kmax_gridtype} highlights the impact of the mesh assignment scheme for increasing order. The assignment scheme tends to smooth the density field for large assignment scheme orders and thus impacts the $\sigg$ value. We also note an impact on the bias $\bsi$. However, no clear trend appears considering the error bars. For the TSC and PCS schemes, the $\fsi$ value is highly biased, even for this baseline mesh configuration, which generally gives large error bars. As confirmed by the more extensive set of fits in figures~\ref{fig:density_variation} and \ref{fig:density_variation_2}, only the NGP scheme gives unbiased $\fsi$ values with reasonable error bars. Considering the previously mentioned large correlation between $\fsi$ and $\sigg$ parameters, we verify that fixing the $\sigg$ value to the one obtained from the NGP fit is not sufficient to remove the bias seen for large-order assignment schemes (TSC, PCS). From now on, we keep only the NGP assignment scheme and use the average of the unbiased NGP fits to fix the $\sigg$ value for the full fit to $6.0\ \mpc$.

Finally, we test the $\kmax$ value impact and mesh shape (cubic or spherical) on the density fit results in figure~\ref{fig:density_kind_kmax_gridtype} (right). The $\kmax$ parameter was extensively discussed in previous studies~\cite{adams_joint_2020,lai_using_2022,carreres_growth-rate_2023}. A large $\kmax$ value implies an explicit trust in small-scale power spectrum modeling. However, in the case of likelihood-based field-level inference, the power spectrum is highly damped at small scales by the mesh window function in equation~\ref{eq:gridding_cor}. Thus, the previous assumption is relaxed when dealing with a large voxel mesh. Equivalently, the assumption on probed scales is entirely driven by the voxel size parameter. The $\kmax$ value is only a criterion for the numerical convergence of the covariance computation. Previous studies~\cite{adams_improving_2017,adams_joint_2020,lai_using_2022} chose a value $\kmax = 0.2$~\hpmpc~and fitted the small-scale range with an additional bias parameter $b_{\mathrm{add}}$. Here, like in \textbf{C23}, we aim to directly use $\kmax = 1.0$~\hpmpc~and skip the addition of this term. The fit results show that both $\kmax$ values give unbiased $\fsi$ values. However, some configurations in figures~\ref{fig:density_variation} and~\ref{fig:density_variation_2} show more instability when the $\kmax = 0.2$~\hpmpc~value is used. As expected, this value is insufficient for small voxel sizes, e.g. $20\ \mpc$, because we do not account for the small scales. As we do not see any clear bias with the $\kmax = 1.0$~\hpmpc~value, we use it for our baseline fit. The previous assumption will not be needed when considering non-linear models in future studies. 

Figure~\ref{fig:density_kind_kmax_gridtype} also shows the impact of a cubic and spherical cut of the mesh. No clear difference is seen in the shown example, except for the decrease in the error bar for the cubic configuration. The latter is expected as the volume probed in the cubic case is larger. We noticed on a more extended fit sample that the cubic configuration tends to give more biased results, and since it is not in agreement with the supernovae cut or the way surveys are built, we keep a spherical cut for the density mesh.

To conclude, this study allowed us to reduce the number of configurations to explore in the following fits by fixing the mesh assignment scheme (NGP), the maximal integration wavenumber ($\kmax = 1.0\ \mpc$), and the shape (spherical). We also fix the value $\sigg = 6.0\ \mpc$ for density-velocity combined fits.

\subsection{ZTF SN Ia velocity field fit}
\label{subsec:velocity}

\begin{figure*}
    \includegraphics[width=0.95\textwidth]{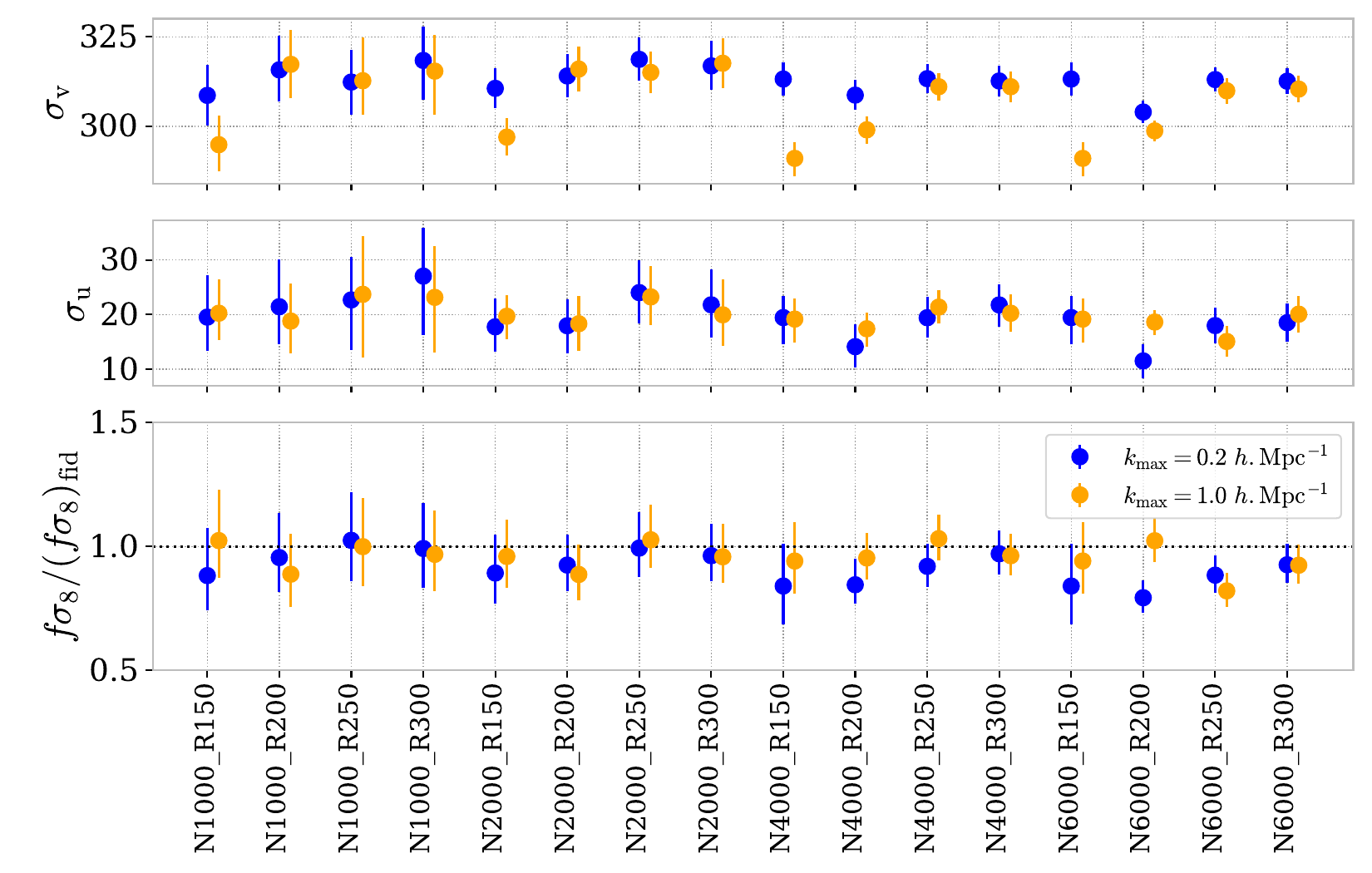}
    \caption{Velocity fit on an idealized \sn~mock using the true simulated velocities, varying the number of supernovae (N) and the maximal comoving distance radius considered (R). The colors show the variation taking two different values of maximal integration wavenumber. As for density, the points are the weighted average of \texttt{iminuit} fit for the 27 mocks, and the error bars are the average \texttt{MINOS} error.}
    \label{fig:velocity_variation}
\end{figure*}

\begin{figure}
    \centering
	\includegraphics[width=0.49\columnwidth]{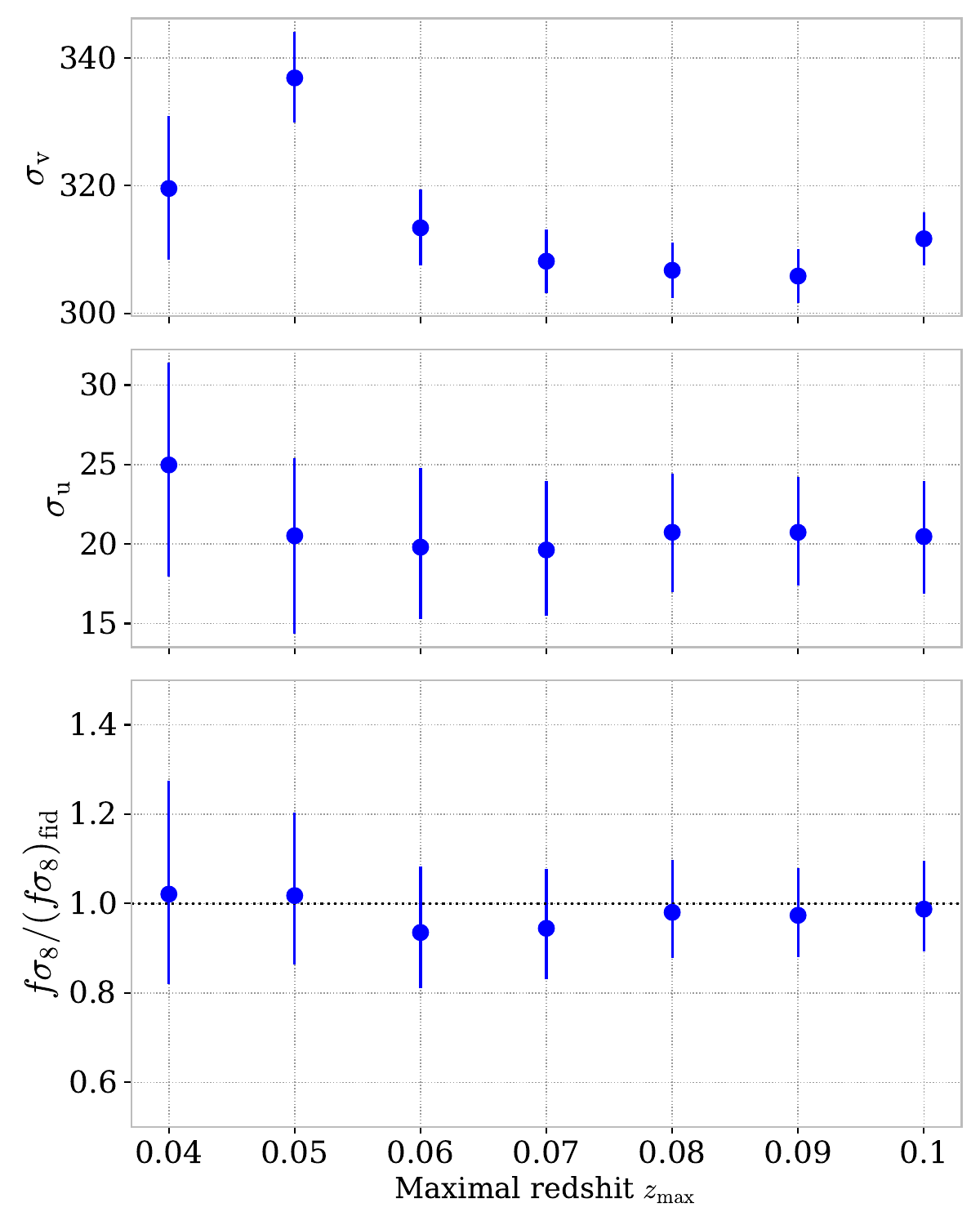}
	\includegraphics[width=0.49\columnwidth]{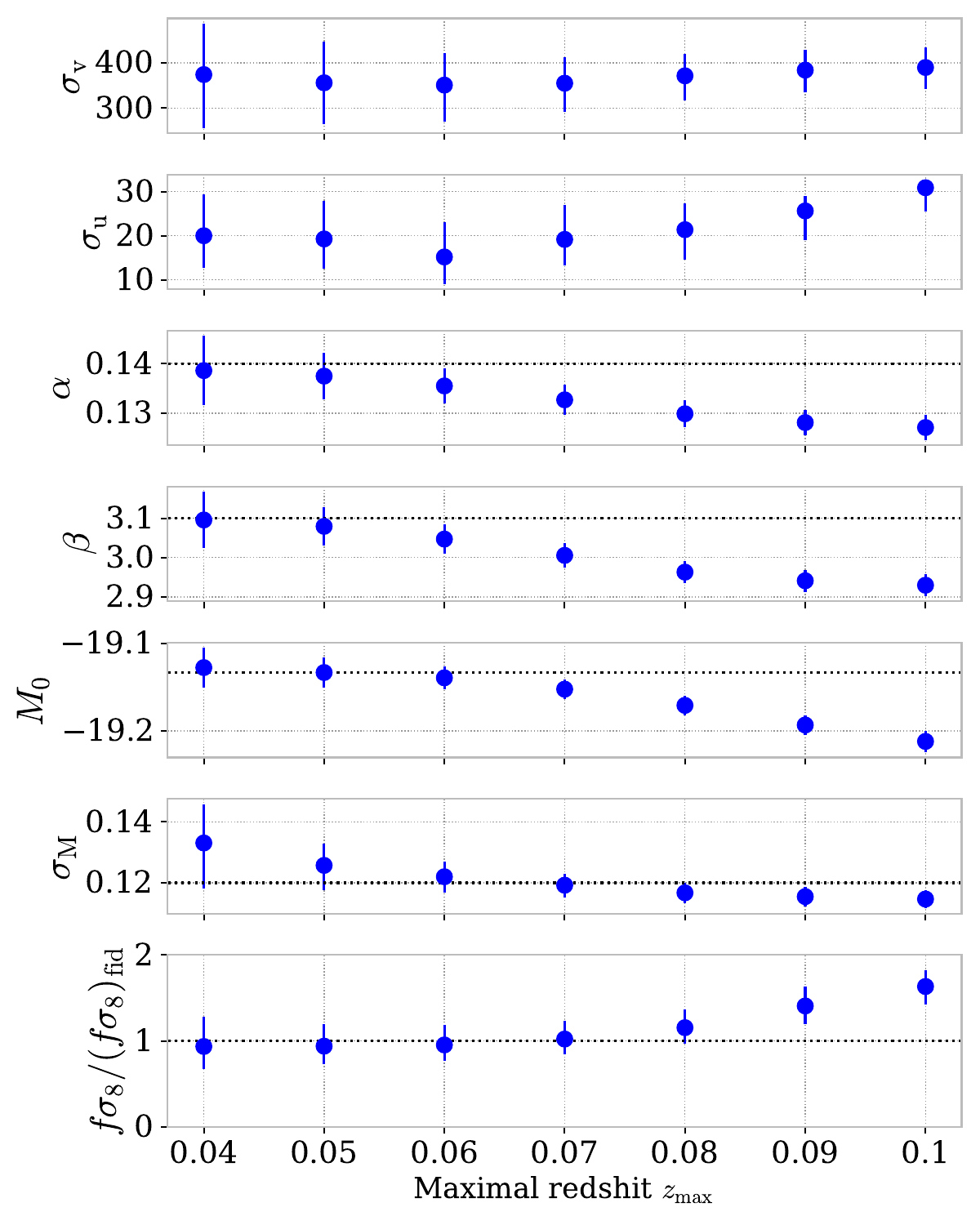}
    \caption{Fit performed with \texttt{iminuit} on the ZTF SN Ia simulation considering the true velocities (left) or the estimated velocities from the Hubble diagram residuals (right). The fit is performed while varying the $\sigu$ value with a Gaussian prior added to the likelihood. As in figure~\ref{fig:velocity_variation}, each point is the weighted average fitted value of the 27 mock realizations, and the error bars are the average \texttt{MINOS} error. The input parameters of the simulation are shown with a dashed black line.}
    \label{fig:velocity_fit_ztflike}
\end{figure}

This section aims to reproduce~\textbf{C23} \sn~velocity fit results on our \texttt{AbacusSummit} simulations. After assessing our \sn~sample in this section, we use it as a reference to measure the gain from adding the density field in the next section. Our velocity model is given by the last term of equation~\ref{eq:model} and is mathematically equivalent to the~\textbf{C23} model.

As for density, we test the velocity fit with different configurations, varying the number of supernovae (1000, 2000, 4000, or 6000), the maximal comoving distance (150, 200, 250, or 300~\hpmpc), and the maximal wavenumber used in the integration (0.2 or 1.0~\hpmpc). Those tests are performed for each mock on the true velocity field, without considering the complete ZTF geometry (only a $\mathrm{DEC} > -30$° cut), and with no prior knowledge of parameters. We aim to detect any dependence of the $\fsi$ parameter on \sn~density in the ideal case, and to verify that a proper configuration is used for the main fit. The results from this test are shown in figure~\ref{fig:velocity_variation}.

For most of the configurations, only the $\kmax = 0.2$~\hpmpc~shows some significant biased values for $\fsi$. This result also indicates that small-scale modes are not adequately accounted for in this case. Since no clear bias is visible for the $\kmax = 1.0$~\hpmpc~value, we choose it for our baseline as in~\textbf{C23} and in accordance with the density fits. For $\kmax = 1.0$~\hpmpc, there are no apparent signs of bias due to \sn~density. The bias is clearly seen for $\kmax = 0.2$~\hpmpc~in the extreme case of 6000 supernovae with a 150~$\mpc$ comoving radius. In this case, the large \sne~density decreases the average \sne~separation, and the modeling is insufficient to obtain an unbiased result.

We also use this test to measure the real value of $\sigu$ in our simulation. The weighted average of unbiased fits with $\kmax = 1.0$~\hpmpc~gives a value $\sigu = 21.5~\mpc$. This value is larger than that obtained in \textbf{C23} on \texttt{OuterRim} simulations. This difference can be caused by the redshift of the \texttt{AbacusSummit} simulation ($z=0.2$) or by differences in the velocity clustering. In any case, since the $\sigu$ parameter cannot be determined a priori from real data, any prior associated to this parameter should be very large to avoid dependence on a specific set of simulations.

We perform the velocity fit with the geometry and the selection effects detailed in section~\ref{subsec:ztfsim}. Only the maximal considered redshift can be varied in this case. As pointed out in \textbf{C23}, there is an important degeneracy between the $\sigu$ and $\fsi$ parameters. Thus, to limit the spread along this degeneracy and to stabilize the fit, we add a Gaussian prior on $\sigu$ to the likelihood. We center this prior on the previously determined value $\sigu = 21.5\ \mpc$. Since the value of this parameter is quite uncertain, we use a substantial standard deviation of $10.0\ \mpc$. Figure~\ref{fig:velocity_fit_ztflike} shows the results of the fit for the 27 mocks, varying maximal redshift and considering true velocity (left) or velocity estimated from the Hubble diagram parameters (right) with equation~\ref{eq:velestimator}. Figure~\ref{fig:velocity_mcmc} in the appendix shows an MCMC sampling of one specific mock for estimated velocities for a redshift cut $z_{\rm max} = 0.07$.

The conclusions of those fits are very similar to \textbf{C23}. For true velocities, the fits are not biased for all $z_{\rm max}$ values. As for \textbf{C23}, the geometry of the ZTF survey shows that the $\fsi$ uncertainty does not drastically decrease for $z_{\rm max} > 0.07$. We interpret this as being caused by the Malquist bias after  $z> 0.07$, which biases the values of $\fsi$.

For estimated velocities, the MCMC fit in figure~\ref{fig:velocity_mcmc} shows very similar correlations between parameters in comparison to \textbf{C23}. In particular, we see strong correlations between $\fsi$ and $\sigu$, and $\sigma_{\rm M}$ and $\sigma_{\rm v}$. The latter correlation is expected because $\sigma_{\rm M}$ accounts for variations in the error on absolute magnitude and $\sigma_{\rm v}$ on the velocity fields, and both are linked by equation~\ref{eq:velestimator}. Considering the \texttt{iminuit} fit results, the same trend as \textbf{C23} appears: including the high redshift \sn~sample biases the fit due to sample selection effects. This conclusion is valid both for $\fsi$ and the Hubble diagram parameters ($\alpha$, $\beta$, and $M_0$). Only the $\sigma_{\rm M}$ does not suffer from this effect. However, since $\sigma_{\rm M}$ is highly correlated with $\sigma_{\rm v}$, this interpretation is invalid. We note that $\fsi$ starts to be biased for larger $z_{\rm max}$ values than Hubble diagram parameters. This study indicates that we can safely use $z_{\rm max} = 0.07$ for our velocity fit.

\subsection{Combining density and velocity fields}
\label{subsec:fullfit}

\begin{figure}
    \centering
	\includegraphics[width=0.525\columnwidth]{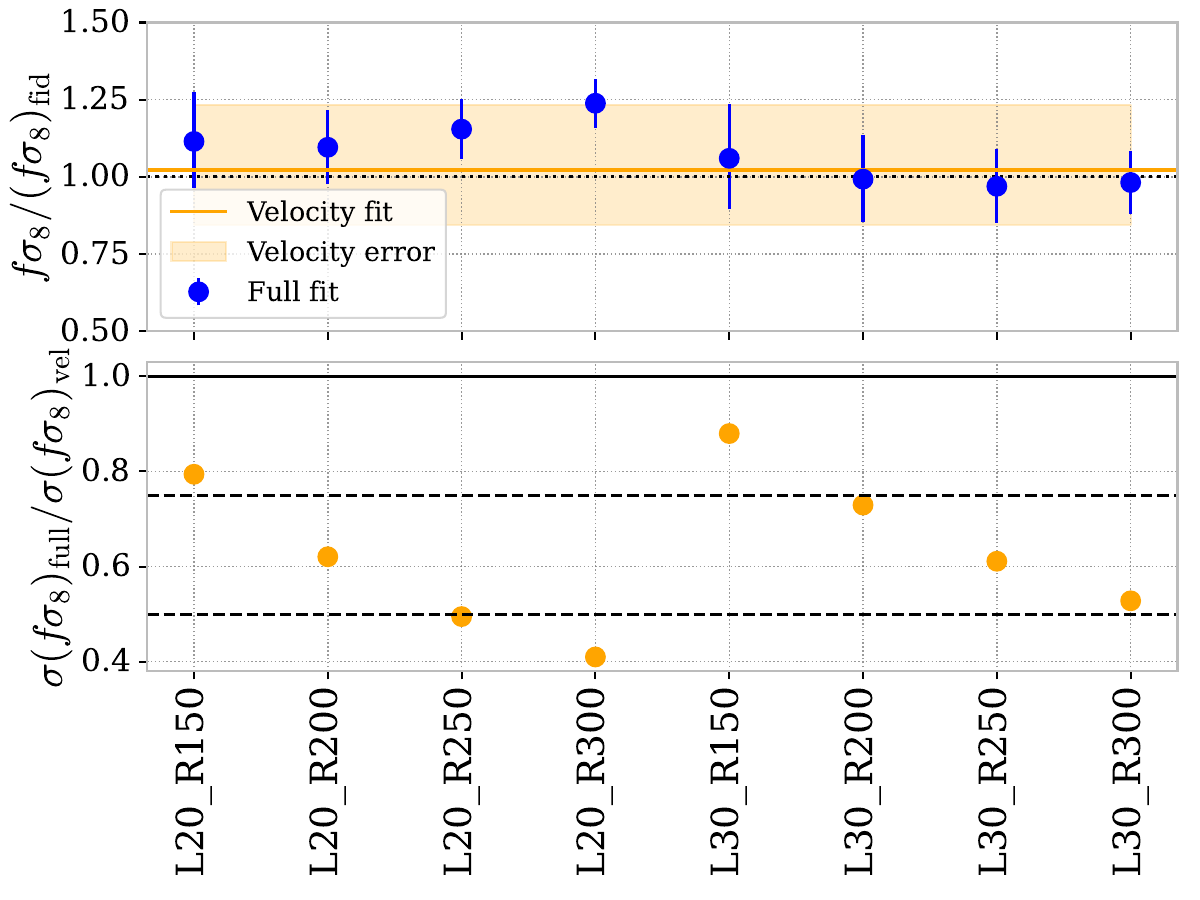}
	\includegraphics[width=0.455\columnwidth]{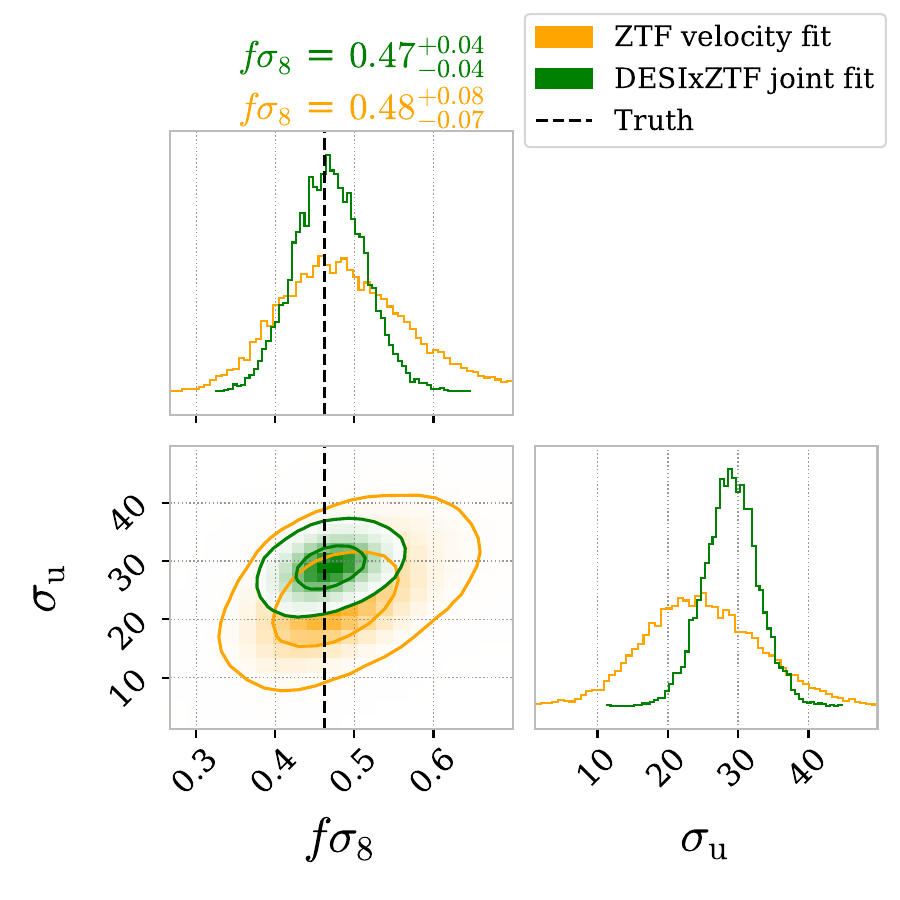}
    \caption{(left) \texttt{iminuit} fit results for the combined fit of ZTF \sn~velocities with DESI BGS density, as a function of the adopted mesh configuration. The values fitted are the weighted average over the \texttt{AbacusSummit} mocks and the uncertainties are the average \texttt{MINOS} errors. The top panel shows the results for the combined fit (blue points) and for the velocity fit (orange line and shaded area). The bottom panel shows the ratio between the combined fit uncertainties and the one obtained with velocity fits. The dashed lines represent a 25 \% and 50 \% improvement. (right) Zoom on the MCMC posterior for the combined fit in green and the velocity fit in orange. Only the contour for $\fsi$ and $\sigu$ is shown. For the density mesh, we take a configuration with the total mesh size of $300\ \mpc$, the voxel size of $30\ \mpc$, a spherical shape, an NGP assignment scheme, and we fix the parameter $\sigg = 6.0\ \mpc$. The black dashed line is the input $\fsi$ value.}
    \label{fig:full_fit}
\end{figure}

Based on the separate conclusions related to BGS galaxy density (section~\ref{subsec:density}) and \sn~velocity fit (section~\ref{subsec:velocity}), we perform the combined fit of both fields using the integration of the full power spectrum model in equation~\ref{eq:model}. Only two configuration parameters are varied for the density mesh: the total mesh size (noted R) and the voxel size (noted L). The maximal redshift for supernovae is fixed to $z_{\rm max} = 0.07$, and the covariance is interpolated with respect to $\sigu$. We note that the $\sigg$ and $\sigu$ parameters could be fitted simultaneously by performing a full two-dimensional interpolation. However, this fit is very expensive numerically as it would require the storage and manipulation of a very large number of covariance matrices. Furthermore, since both parameters aim to model the effect of FoG on the position of each field, those parameters would be highly correlated. Consequently, we fix the value of the density FoG parameter to $\sigg = 6.0\ \mpc$ based on the density results. We note that obtaining the value of $\sigg$ can always be done with the density fit performed first, even for data. However, as the $\sigg$ value is potentially correlated to $\sigu$ by the density-velocity cross terms, we note that fixing $\sigg$ could be a source of systematic bias and will be explored in future studies.

Figure~\ref{fig:full_fit} shows the combined fit for the \texttt{iminuit} minimization and the MCMC sampling. It also exhibits the comparison with the velocity fit for the equivalent configuration. The fit for all parameters is in the appendix in figure~\ref{fig:full_minuit}. The full MCMC posterior for the combined fit is given in figure~\ref{fig:full_mcmc}. The density mesh configuration impacts the $\fsi$ value and decreases the error bar for all cases. As for density fits, this decrease is directly correlated with the number of density mesh voxels. However, for too large values of total mesh size with $L = 20\ \mpc$ voxel size, the $\fsi$ value starts to show a small $1\sigma$ bias. This is caused by the inability of our linear model to perform a correct fit for those separations. The density configuration with $30\ \mpc$ voxel size gives an excellent improvement on the $\fsi$ uncertainty up to 50 \% without introducing any bias. The absence of bias indicates that our linear model is suitable for those separations. If we consider the density configuration with $L = 30\ \mpc$ voxel size and $R = 300\ \mpc$ total size as the conservatively best configuration, the mean bias on $\fsi$ is $1.8$ \% and the mean error bar achievable by a DESI ZTF measurement is $\sigma_{\fsi} = {}^{+0.047}_{-0.048}$ corresponding to a 10 \% error. 

The result of MCMC sampling shows that the combined fit also decreases the error bar of the $\sigu$ nuisance parameter. Considering the error bars, the values are compatible at the 1$\sigma$ level. This decrease is expected from the modeling of the cross-correlation between velocity and density introduced by the $\mathcal{P}_{gv}$ term in equation~\ref{eq:model}. Similarly, the MCMC posteriors show a significant correlation between $\fsi$ and $\bsi$ parameters, as expected given the presence of their product in the $\mathcal{P}_{gv}$ and $\mathcal{P}_{gg}$ terms. As highlighted in figure~\ref{fig:full_minuit}, this correlation does not seem to cause the $\fsi$ bias seen with the $L = 20\ \mpc$ voxel size fits.

To conclude, the full fit shows a decrease in the $\fsi$ error bar for all configurations. It indicates that galaxy density field correlations should always be added to maximize the constraining power for future studies. In the specific case of the DESIxZTF simulation, the gain on $\fsi$ error bar can go up to 50 \% without introducing any bias. Finally, as pointed out by the $L = 20\ \mpc$ voxel size fits, an improvement in small-scale modeling could further increase the gain in constraining power.

\section{Conclusion and prospects}
\label{sec:conclusion}

We have created precise simulations mimicking the clustering of the DESI DR3 BGS galaxy sample and ZTF Y6 \sn~sample on an \texttt{AbacusSummit} simulation. These simulations are used to obtain a precise forecast of the $\fsi$ constraint achievable with the two surveys. As the corresponding real datasets will be available shortly, our simulations will serve as a basis for their analysis. We applied a generalized likelihood-based field-level inference algorithm to fit the density and velocity fields of these simulations.

An in-depth study of the fits alone on the BGS density field allowed us to determine the ideal mesh configurations. We conducted a similar study to that of~\cite{carreres_growth-rate_2023} on the velocity field based on \sne. Reproducing these results on the \texttt{AbacusSummit} simulation allowed us to validate the velocity clustering and highlight some \sn~density dependence effects. Finally, we performed the fit on the combination of galaxy density and velocity field. This study allowed us to highlight the correlations between the parameters related to both fields. We also concluded that adding the DESI BGS DR3 density field to the ZTF \sn~dataset would improve the uncertainty on $\fsi$ by up to 50 \% without introducing any bias.

This study could be extended in order to directly apply it to the DESI and ZTF datasets. From the \sn~point of view, additional systematics should be considered, such as the intrinsic scattering modeling (see~\cite{Carreres2025} for a study of the effect on simulated LSST data). The impact of specific ZTF instrumental effects on the magnitude of the \sne~should also be accounted for, or at least tested, so that it does not introduce bias in $\fsi$. On the density side, an exploration of the impact of target selection and associated photometric systematics should be conducted.

Combining density and velocity fields is a methodology that can be extended to several different types of surveys. On the Legacy Survey of Space and Time (Rubin-LSST)~\citep{ivezic_lsst_2018}, several studies were already performed to estimate $\fsi$ on the \sn~velocity field~\cite{Rosselli2025,Carreres2025}. Those constraints can be improved by combining the velocity field with the galaxy density field from the 4-meter Multi-Object Spectroscopic Telescope (4MOST)~\citep{de_jong_4most_2019}, specifically the Cosmology Redshift Survey (4MOST-CRS)~\citep{richard_4most_2019,Verdier2025} and the Hemisphere Survey of the Nearby Universe (4MOST-4HS)~\citep{taylor_4most_2023}. The statistics of the velocity field can also be largely improved with surveys other than ZTF or LSST. The latest release of the Asteroid Terrestrial Last Alert System (ATLAS) survey~\cite{Marlin2025} will provide a large sample of $3,000$ cosmology graded \sne. Even if it contains some objects in common with ZTF, ATLAS can be used to create a $\fsi$ measurement with separated instrumental systematics, thus improving the robustness of the measurement. The Dark Energy Bedrock All-Sky Supernovae (DEBASS)~\cite{Sherman2025} will provide a smaller sample of \sne~but with improved calibration from spectroscopy. This sample can also be used to perform a $\fsi$ measurement, focusing on the hypothesis linked to velocity estimation. On the longer term, the La Silla Schmidt Southern Survey (LS4)~\cite{Miller2025} will provide a supernova sample suitable for peculiar velocity studies in the southern sky. 

The \sn~velocities can also be coupled in our framework to the velocities estimated from galaxy surveys with TF and FP with, e.g., the use of the DESI peculiar velocity samples~\citep{Saulder2023,Said2024}. The coupling of all the future datasets available (DESI and 4MOST density fields, DESI peculiar velocities, and \sne~from ZTF, LSST, and LS4) will lead to the most precise measurement of $\fsi$ at low redshifts. 

Performing those large analysis will require major modeling improvements. The velocity likelihood can be modified to account for the sample selection effect seen on velocities in the right panel of figure~\ref{fig:velocity_fit_ztflike}. Our field-level inference method should be extended to account for different types of velocity fields with different estimators and characteristic uncertainties. As pointed out by our analysis, the small-scale modeling can be largely improved by switching to a non-linear model with, e.g., the Effective Field Theory of Large Scale Structure (EFTofLSS)~\cite{Carrasco2013,Perko2016,DAmico2020,Ivanov2022}. Since many EFTofLSS parameters will not be present in the linear decomposition of the power spectrum model in equation~\ref{eq:Pwflip}, the implementation of those non-linear models will need more flexible regression methods such as Gaussian processes or neural networks. Finally, our field-level inference method is limited to the $\fsi$ cosmological parameter but can be extended to fit for additional parameters such as $H_0$~\cite{Piras2025}, a cosmological dipole, or a redshift-dependent growth model ($\Omega_{\rm m}, \gamma$) such as in~\cite{Crisman2026}.

\appendix

\section{Complementary figures}
\label{appendix:full_plots}

\begin{figure}
	\includegraphics[width=\textwidth]{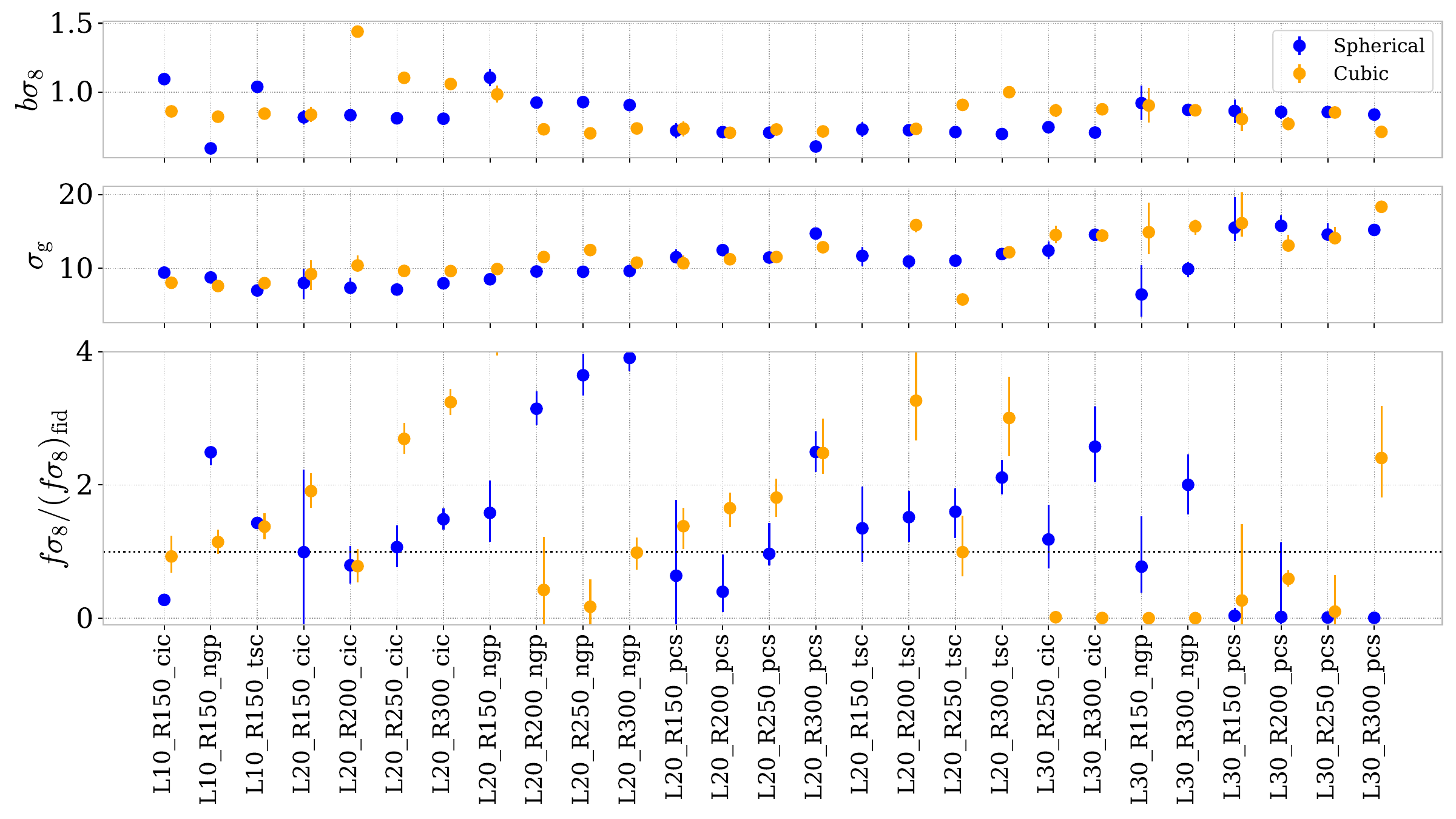}
    \caption{Fit of the DESI BGS density field with DESI footprint for all the mesh configurations which give valid results and with a size smaller than $15,000$ elements. The results are shown for a spherical (blue points) or cubic (orange points) mesh configuration with a maximal wavenumber integration of $k_{\rm max} = 0.2$ \hpmpc. The blue dots are the weighted average fitted values of the 27 BGS mocks, and the error bars are the average \texttt{MINOS} errors. The labels give the voxel size (L), the total mesh size (R), and the assignment scheme (ngp, cic, tsc, or pcs).}
    \label{fig:density_variation}
\end{figure}

\begin{figure}
	\includegraphics[width=\textwidth]{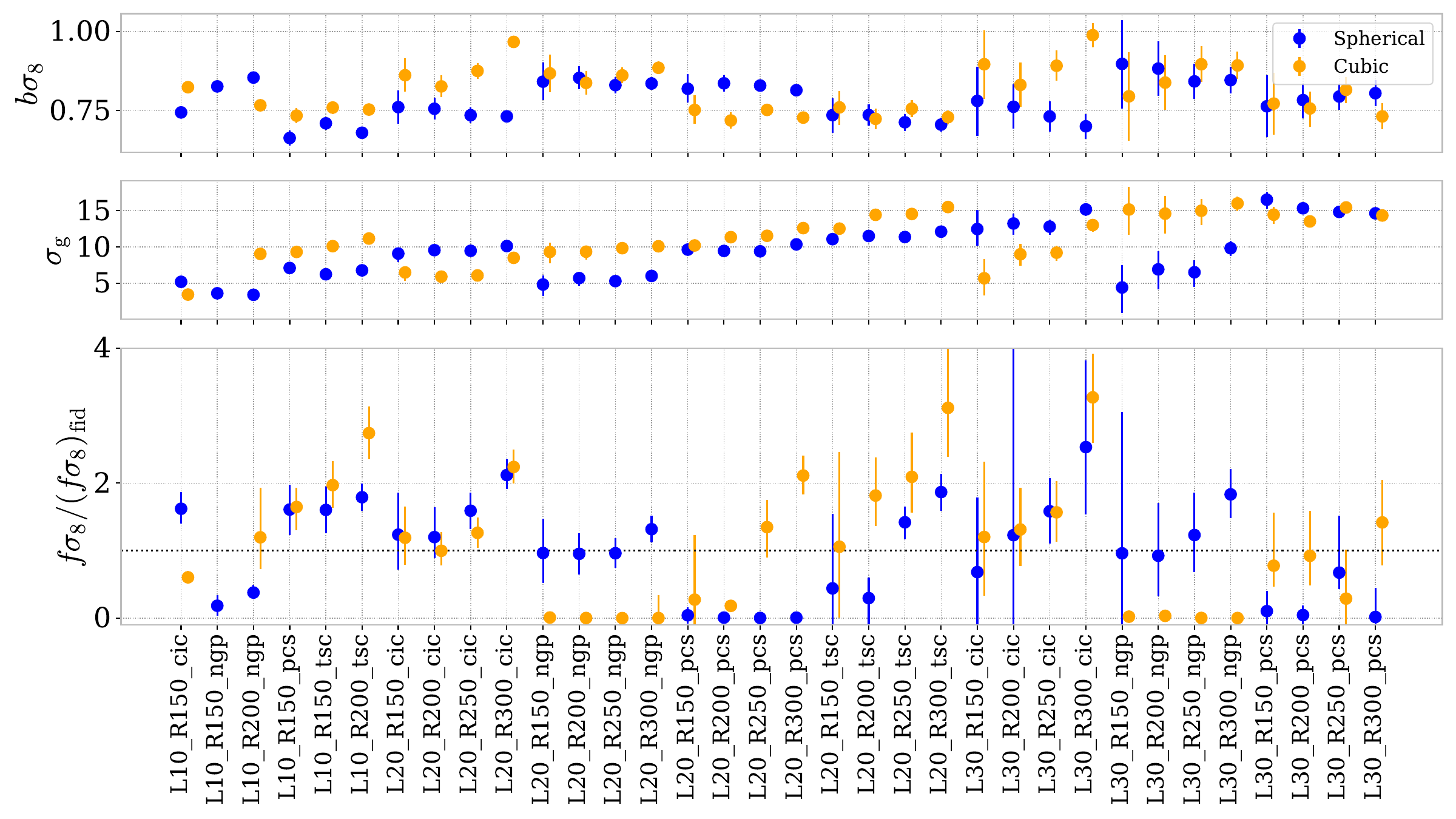}
    \caption{Same as figure~\ref{fig:density_variation} with maximal integration wavenumber $k_{\rm max} = 1.0$ \hpmpc.}
    \label{fig:density_variation_2}
\end{figure}

\begin{figure}
	\includegraphics[width=\textwidth]{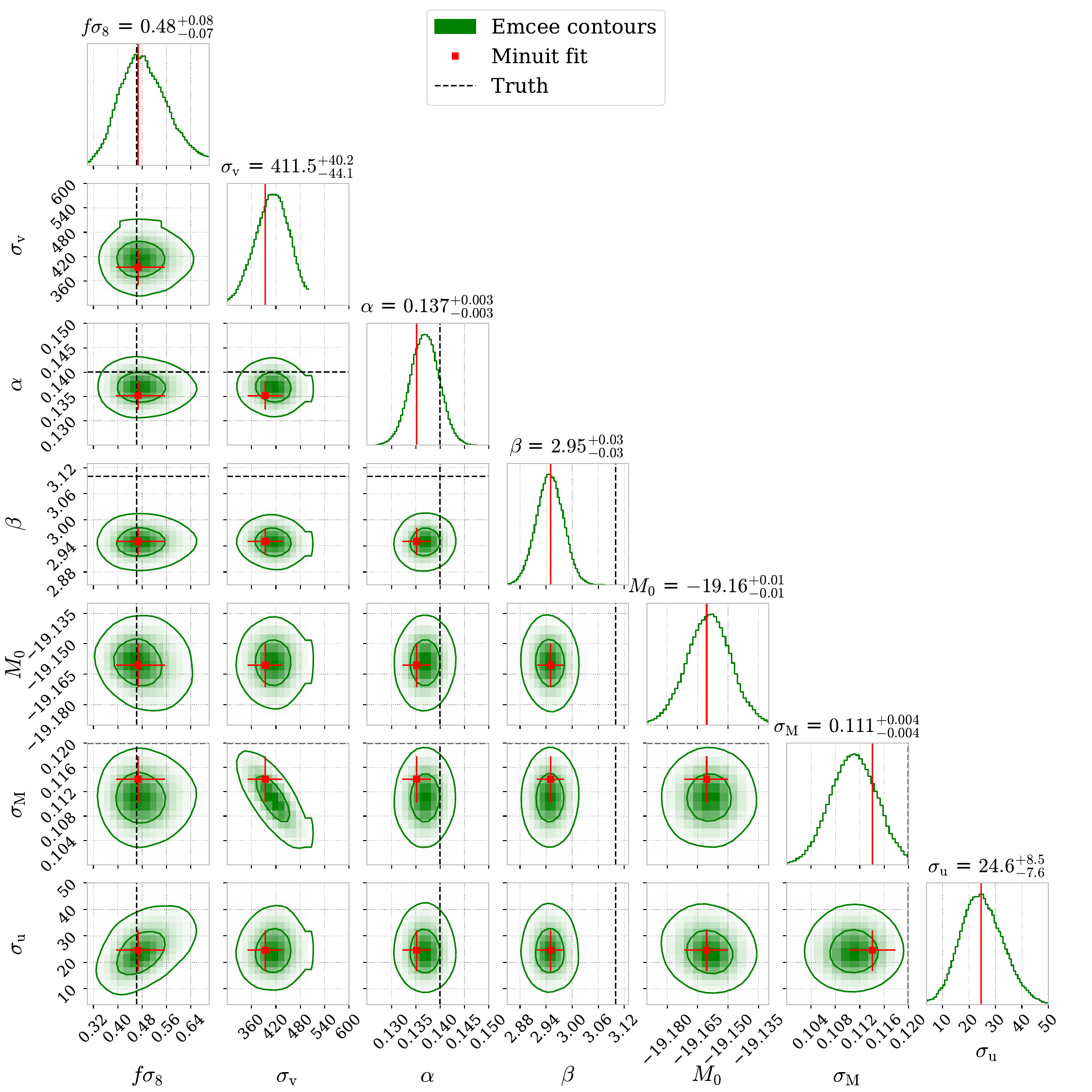}
    \caption{Markov chain Monte Carlo sampling of the velocity likelihood of one ZTF Y6 mock. The velocities are estimated from the Hubble diagram residuals and the standardization of the Hubble diagram is also sampled ($\alpha, \beta, M_0, \sigma_{\rm M}$). The green contours are the 1- and 2-$\sigma$ MCMC contours. The red points are the \texttt{iminuit} fit and \texttt{MINOS} errors. The dashed black line shows the input parameters. The maximal redshift taken is $z_{\rm max} = 0.07$.}
    \label{fig:velocity_mcmc}
\end{figure}

\begin{figure}
	\includegraphics[width=\textwidth]{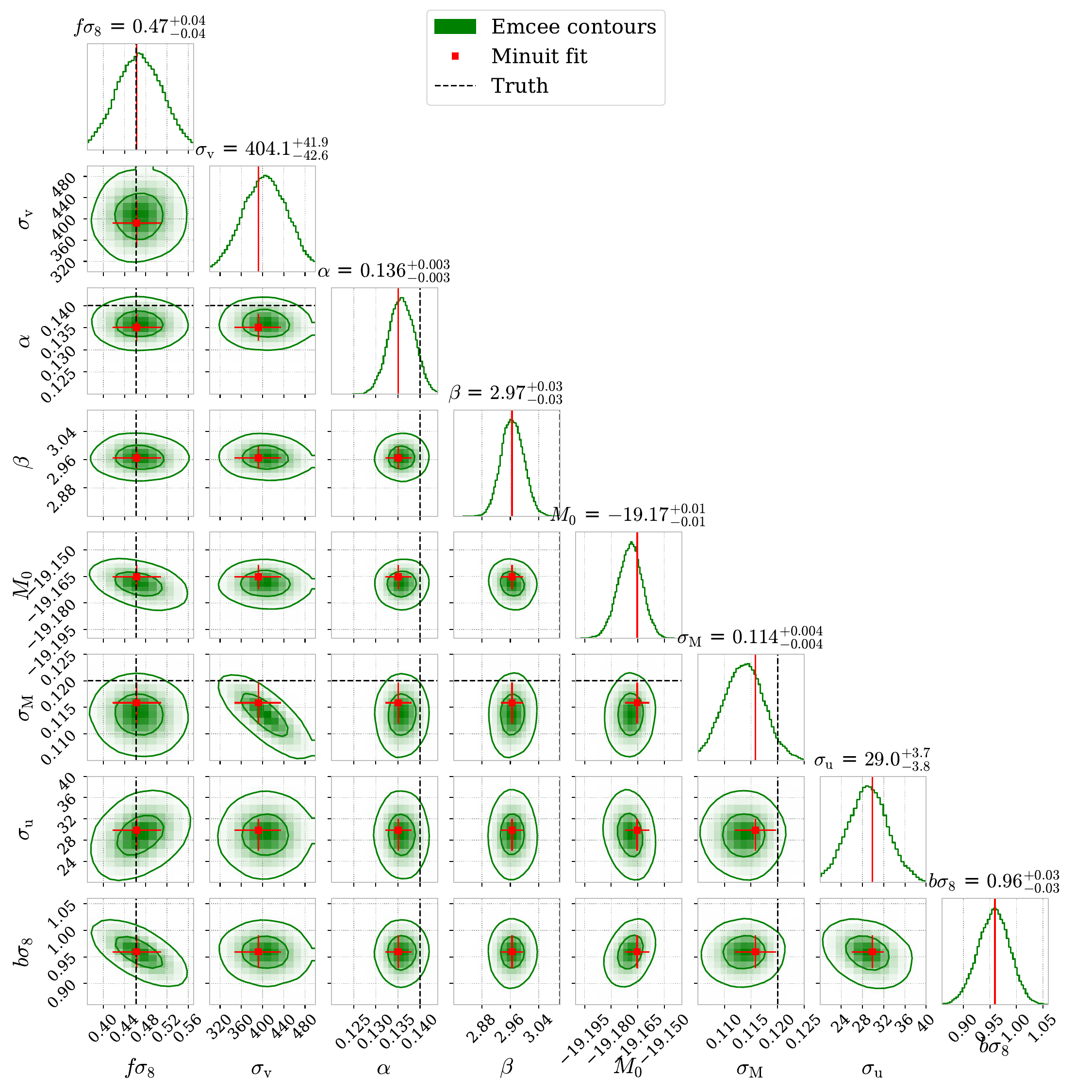}
    \caption{Same as figure~\ref{fig:velocity_mcmc} but on the full likelihood containing ZTF Y6 estimated velocities and DESI BGS Y5 galaxy density field. Only the $\bsi$ parameter is added. For the density mesh, we take a configuration with the total mesh size of $300\ \mpc$, the voxel size of $30\ \mpc$, a spherical shape, and an NGP assignment scheme. We fix the parameter $\sigg = 6.0\ \mpc$.}
    \label{fig:full_mcmc}
\end{figure}

\begin{figure}
    \centering
	\includegraphics[width=0.7\textwidth]{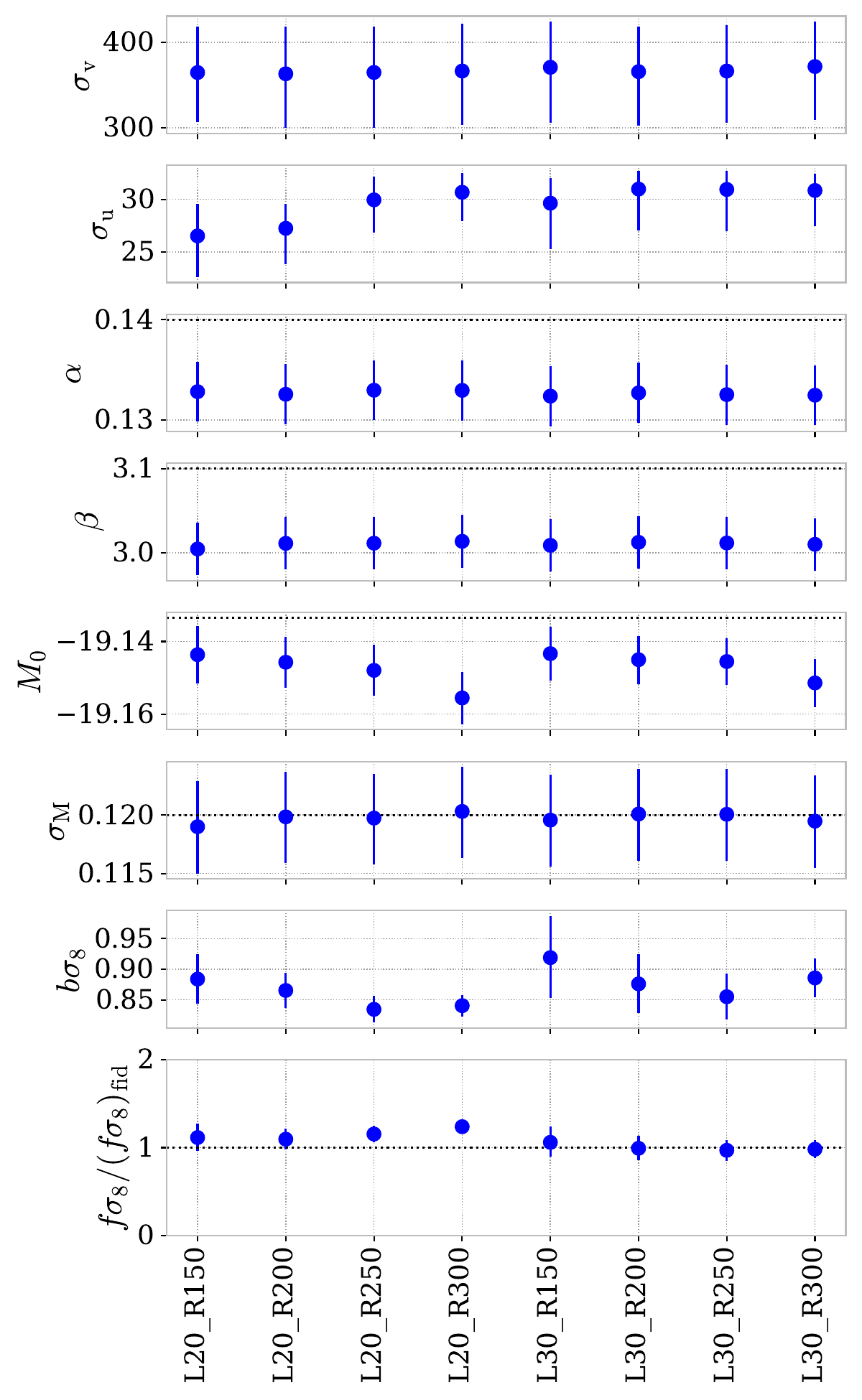}
    \caption{Best fit with \texttt{iminuit} of the full likelihood containing ZTF Y6 estimated velocities and DESI BGS Y5 galaxy density field. The values are the weighted average of the 27 mocks, and the error bar are the average \texttt{MINOS} errors. Density mesh configuration is varied with respect to total mesh size (R) and voxel size (L). The dashed black line shows the input parameters.}
    \label{fig:full_minuit}
\end{figure}

This appendix contains additional figures in order to complement the results shown in the main text. 

Figures~\ref{fig:density_variation} and \ref{fig:density_variation_2} show all the valid \texttt{iminuit} density fits for all the spherical mesh configurations and $\kmax$ variation tested to complement the results from section~\ref{subsec:density}. As the cubic and spherical configurations give very similar results, the cubic one is not shown for clarity. For clarity, the results for maximal wavenumber are separated, with $\kmax = 0.2~\mpc$ in figure~\ref{fig:density_variation} and $\kmax = 1.0~\mpc$ in figure~\ref{fig:density_variation_2}. The figures indicated that only the NGP mesh assignment scheme is giving reasonable $\fsi$ values and that the $\kmax = 1.0~\mpc$ maximal wavenumber is more stable.

Figure~\ref{fig:velocity_mcmc} represents the \texttt{emcee} posterior contour obtained from the sampling of the ZTF \sn~velocity likelihood. A similar posterior for the combined fit is given in figure~\ref{fig:full_mcmc}, adding the DESI BGS density field and thus the $\bsi$ parameter. The ZTF velocity MCMC posterior is giving the same correlation profiles as \textbf{C23}. In addition, the combined fit in figure~\ref{fig:full_mcmc} shows a clear correlation between the $\bsi$ and $\fsi$ parameters, highlighting the need for a precise characterization of the density field as a first step. Finally, figure~\ref{fig:full_minuit} shows the details for the \texttt{iminuit} combined fits for all the configurations tested with a maximal redshift for \sn~of $z_{\rm max} = 0.07$.

\acknowledgments

This material is based upon work supported by the U.S. Department of Energy (DOE), Office of Science, Office of High-Energy Physics, under Contract No. DE–AC02–05CH11231, and by the National Energy Research Scientific Computing Center, a DOE Office of Science User Facility under the same contract. Additional support for DESI was provided by the U.S. National Science Foundation (NSF), Division of Astronomical Sciences under Contract No. AST-0950945 to the NSF’s National Optical-Infrared Astronomy Research Laboratory; the Science and Technology Facilities Council of the United Kingdom; the Gordon and Betty Moore Foundation; the Heising-Simons Foundation; the French Alternative Energies and Atomic Energy Commission (CEA); the Secretariat of Science, Humanities, Technology and Innovation (SECIHTI) of Mexico; the Ministry of Science, Innovation and Universities of Spain (MICIU/AEI/10.13039/501100011033), and by the DESI Member Institutions: \url{https://www.desi.lbl.gov/collaborating-institutions}. Any opinions, findings, and conclusions or recommendations expressed in this material are those of the author(s) and do not necessarily reflect the views of the U. S. National Science Foundation, the U. S. Department of Energy, or any of the listed funding agencies.

The authors are honored to be permitted to conduct scientific research on I'oligam Du'ag (Kitt Peak), a mountain with particular significance to the Tohono O’odham Nation.

ZTF Simulation logs are based on observations obtained with the Samuel Oschin Telescope 48-inch and the 60-inch Telescope at the Palomar Observatory as part of the Zwicky Transient Facility project. ZTF is supported by the National Science Foundation under Grants No. AST-1440341 and AST-2034437 and a collaboration including current partners Caltech, IPAC, the Weizmann Institute of Science, the Oskar Klein Center at Stockholm University, the University of Maryland, Deutsches Elektronen-Synchrotron and Humboldt University, the TANGO Consortium of Taiwan, the University of Wisconsin at Milwaukee, Trinity College Dublin, Lawrence Livermore National Laboratories, IN2P3, University of Warwick, Ruhr University Bochum, Northwestern University and former partners the University of Washington, Los Alamos National Laboratories, and Lawrence Berkeley National Laboratories. Operations are conducted by COO, IPAC, and UW.

The project leading to this publication has received funding from Excellence Initiative of Aix-Marseille University - A*MIDEX, a French ``Investissements d'Avenir'' program (AMX-20-CE-02 - DARKUNI).

\paragraph{Data availability} 
\ 

All the figure data points in this article are made available according to the data management policy of DESI\footnote{\label{zenodo_data_release}\url{https://doi.org/10.5281/zenodo.20446051}}. All the fits and plots of this article are generated with \texttt{flip}$^{\ref{corentinravoux/flip}}$ (1.0.0). 

\bibliographystyle{JHEP}
\bibliography{biblio}

\end{document}